\documentclass[aps,prl,twocolumn,superscriptaddress,floatfix,longbibliography]{revtex4-2}
\usepackage{graphicx,amsmath,amssymb,epstopdf,wasysym}
\usepackage[space]{grffile}
\usepackage[dvipsnames]{xcolor}
\usepackage{braket,amsfonts}
\usepackage{dsfont}
\usepackage{float}
\usepackage{cancel}
\usepackage{comment}
\usepackage[normalem]{ulem}

\usepackage{hyperref}
    \hypersetup
    {   
        unicode=true,           
        pdftoolbar=true,        
        pdfmenubar=true,        
        pdffitwindow=false,     
        pdfstartview={FitH},    
        pdfauthor={Ivan Pasqua, Carlos Mejuto-Zaera, Michele Fabrizio},
        pdftitle={Fractionalized Fermi liquids with the ghost-Gutzwiller Ansatz},
        pdfkeywords={},
        pdfnewwindow=true,      
        colorlinks=true,        
        linkcolor=MidnightBlue,        
        citecolor=Maroon,        
        filecolor=Brown,        
        urlcolor=Maroon          
    }

\newcommand{\up}{\uparrow}
\newcommand{\down}{\downarrow}
\newcommand{\be}{\begin{equation}}
\newcommand{\ee}{\end{equation}}
\newcommand{\bk}{\mathbf{k}}
\newcommand{\bR}{\mathbf{R}}
\newenvironment{eqs}%
{\begin{equation} \begin{aligned}}%
{\end{aligned} \end{equation} }
\newcommand{\beal}{\begin{eqs}}
\newcommand{\eal}{\end{eqs}}
\newcommand{\eqn}[1]{(\ref{#1})}

\newcommand{\dagga}{{\phantom{\dagger}}}

\newcommand{\fract}[2]{\frac{\displaystyle #1}{\displaystyle #2}}
\newcommand{\bw}{\begin{widetext}}
\newcommand{\ew}{\end{widetext}}
\newcommand{\esp}[1]{\text{e}^{#1}}

\newcommand{\bd}[1]{{\boldsymbol{#1}}}

\newcommand{\Tr}{\mathrm{Tr}}
\footnotetext{Contact author: ipasqua@sissa.it}

\begin{document}

\title{Fractionalized Fermi liquids with the ghost-Gutzwiller Ansatz}

\author{Ivan Pasqua$^\ast$}
\affiliation{International School for Advanced Studies (SISSA), Via Bonomea 265, I-34136 Trieste, Italy}

\author{Carlos Mejuto-Zaera}
\affiliation{Univ Toulouse, CNRS, Laboratoire de Physique Th\'{e}orique, Toulouse, France}

\author{Michele Fabrizio}
\affiliation{International School for Advanced Studies (SISSA), Via Bonomea 265, I-34136 Trieste, Italy}

\date{\today}

\begin{abstract}
Fractionalized Fermi liquids ($\mathrm{FL}^{\!*}$), elusive metallic states characterized by fractionalized quasiparticles alongside conventional ones and defying Luttinger's theorem, are prime candidates for the pseudogap regime of underdoped cuprates. We show that a $\mathrm{FL}^{\!*}$ phase emerges in the single-band $t$-$J$ model through a simple ghost-Gutzwiller Ansatz, optimized at the cost of a mean-field calculation. The resulting temperature-doping phase diagram encompasses a low-doping $\mathrm{FL}^{\!*}$, a $d$-wave superconducting dome, and an overdoped conventional Fermi liquid, thereby reproducing key qualitative features of cuprate phenomenology.
\end{abstract}

\maketitle

\textit{Introduction.}---The pseudogap normal phase in the underdoped regime is one of the main puzzles surrounding high-T$_c$ cuprate superconductors \cite{Norman1998,DoironLeyraud2007,Badoux2016,Keimer2015}. A longstanding and controversial question, aside from the mechanism behind the pseudogap opening at the anti-nodal points \cite{Andrei-NatCom2025}, is the topology of the residual Fermi surface at the nodal points.  Recent angle-dependent magnetoresistance measurements \cite{Fang-NatPhys2022,Chan-NatPhys2025} have finally provided strong and potentially conclusive evidence for the existence of small Fermi pockets that account just for the doping away from half-filling rather than the total number of conduction electrons. This clear violation of  Luttinger's theorem \cite{Luttinger} is compatible with a pseudogap originating from a Luttinger surface that appears in the Brillouin zone \cite{YangRiceZhang2006,Heath_2020,Jan-PRB2022} upon approaching the half-filled Mott insulator. An appealing potential consequence of this scenario is the coexistence of conventional quasiparticles hosted by the Fermi pockets and neutral but spinful ones \cite{mio-2}, much in the spirit of Anderson's spinons in the RVB theory of weakly doped antiferromagnetic Mott insulators \cite{Anderson1987,BaskaranZouAnderson1987}.\\
\noindent
A popular and insightful framework to describe the spin-charge fractionalization of the physical electron \cite{LeeNagaosaWen2006} is the parton construction, which represents the electron as the product of a boson and a fermion auxiliary particles, one of which carries the spin and the other the charge   
\cite{BaskaranZouAnderson1987,LiuKotliar1988,Wen-PRL2005,Ng-PRB2005,Punk-PNAS2015,Georges-PNAS2018,Georges-PRX2018,Punk-PRR2020,Nyhegn-PRXQ2025}. 
In this approach, the quasiparticles at the Fermi pockets, which carry both charge and spin, would appear as bound states of the auxiliary particles. \\
\noindent
Alternatively, one might wish to represent the physical electrons directly in terms of distinct fermionic quasiparticles: one neutral and spinful and the others bearing both charge and spin. 
A compelling theoretical realization of this strategy is 
provided by \textit{fractionalized Fermi liquids} ($\mathrm{FL}^{\!*}$) \cite{SenthilSachdevVojta2003,SenthilVojtaSachdev2004}, whose hallmark is precisely the violation of Luttinger's theorem. $\mathrm{FL}^{\!*}$ therefore stands out as a natural theoretical target for the pseudogap metal. The $\mathrm{FL}^{\!*}$ scenario was originally formulated in Kondo-lattice models where itinerant carriers and localized spins are inherently part of the model. Extending this to single-band models representative of doped Mott insulators is not immediately obvious. A promising attempt in this direction is the recently developed ancilla-qubit construction \cite{ZhangSachdev2020,Zhou-PRB2025,Bonetti_2026}. This approach maps the original single-band problem onto an auxiliary ancilla model with hidden spin layers coupled among themselves and to the physical layer via Kondo exchange terms.\\
\noindent
The concept of adding auxiliary fermions beyond the physical electrons to improve the variational description of correlated systems also underpins the so-called ghost-Gutzwiller (gGut) approximation \cite{Lanata2017}. 
This method is attracting growing attention as a reliable, versatile and computationally tractable approach to strongly correlated electrons~\cite{guerci2019,frank2021,Lee2023a,Lee2023b,Mejuto2023a,Tagliente2025,Pasqua2026,Mejuto2026,ArXivSamuele2026a,ArXivSamuele2026b,ArXivMoridani2026}. \\
\noindent
In this Letter we demonstrate that the ghost-Gutzwiller approximation can straightforwardly stabilize a $\mathrm{FL}^{\!*}$ phase in the $t$-$J$ model for cuprate superconductors \cite{ZhangRice1988}. To achieve this we will deliberately employ a simple and physically motivated variational wavefunction whose optimization demands the same computational effort as a mean-field calculation. The resulting variational phase diagram, encompassing $\mathrm{FL}^{\!*}$, superconducting and conventional Fermi liquid (FL) phases, has a remarkable qualitative similarity to the phenomenology of high-T$_c$ superconductors.\\      
\begin{figure}[tbp]
    \centering
    \includegraphics[width=0.5\textwidth]{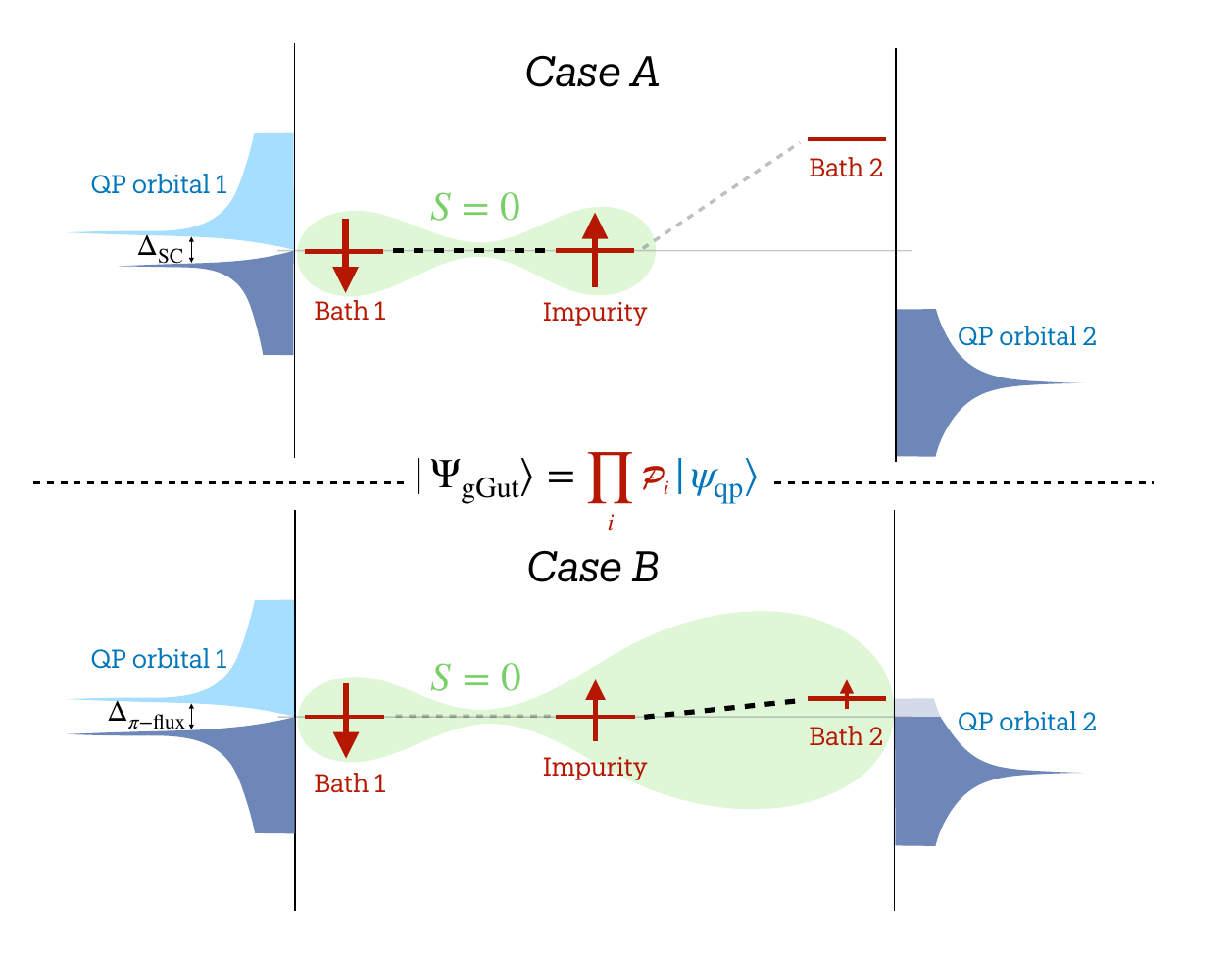}
    \caption{Schematic structure of the two-bath ghost-Gutzwiller Ansatz in the
    two cases considered in the main text. In each panel the central cartoon
    depicts the impurity model, while the blue lobes on either side are the densities of states of the corresponding auxiliary quasiparticle (QP) bands. Green shading marks a spin singlet ($S=0$) and a black (gray) dashed line a non-vanishing (vanishing) hybridization. In case A~\eqref{psi-A}, the impurity hybridizes and forms a singlet with bath 1, which accommodates the doped charge; the associated auxiliary QP band 1 may develop $d$-wave quasiparticle pairing $\Delta_{\mathrm{SC}}$. Bath~2 stays empty and
    decoupled. In case B~\eqref{psi-B}, bath 1 remains half-filled and
    unhybridized, but spin-entangled with the impurity, so that QP orbital 1 represents a dispersive spinon (no physical electron spectral weight, $R_1=0$) that may develop $\pi$-flux pairing $\Delta_{\pi\text{-flux}}$, while bath 2 is promoted to the Fermi level and carries the doped charge.}
    \label{fig:gGutWF}
\end{figure}
\textit{Model and variational wavefunction.}---The Hamiltonian of the $t$-$J$ model reads
\begin{equation}
H = - \sum_{i j \sigma}\, t_{ij} \left( c^\dagger_{i \sigma} c^\dagga_{j \sigma} + \mathrm{H.c.} \right) + J \sum_{\langle ij \rangle}\,\boldsymbol{S}_i \cdot \boldsymbol{S}_j\,,
\label{Ham}
\end{equation}
with $t_{ij}$ the hopping amplitude between sites $i$ and $j$ of a square lattice, $J>0$ the exchange coupling between nearest-neighbor spins $\boldsymbol{S}_i = \sum_{\alpha \beta}\, c^\dagger_{i \alpha}\,  \bd{S}_{\alpha \beta}\, c^\dagga_{i \beta}$, $\bd{S}$ the vector of spin-1/2 matrices, and $c^\dagger_{i \sigma}$ the creation operator for an electron with spin $\sigma$ at site $i$. The local Hilbert space is constrained by the no-double occupancy condition, and we fix the average occupation number per site to $n = 1 - \delta$, with $\delta>0$.\\
The Gutzwiller wavefunction is defined as \cite{Gutzwiller1963,Gutzwiller1965,Bunemann1998,Fabrizio2007,Yao2014,Yao2015,lanata2015,fabrizio2017,Lanata2017}  
\begin{equation}
    \ket{\Psi_\mathrm{gGut}} = \prod_{i}\, \mathcal{P}_i\, \ket{\psi_\mathrm{qp}}\,,\label{G-wavefunction}
\end{equation}
where the uncorrelated $\ket{\psi_\mathrm{qp}}$ is, by construction, a Fock state, thus a Slater determinant or BCS wavefunction, so as to leverage Wick's theorem, and $\mathcal{P}_i$ a local linear operator. The parameters defining the wavefunction \eqn{G-wavefunction} are optimized by minimizing the ground-state energy within the Gutzwiller approximation (GA), which becomes exact in the limit of infinite lattice coordination. The local operator $\mathcal{P}_i$, which can be chosen to be site-independent under the assumption of site-independent local single-particle density matrix, $\mathcal{P}_i \equiv \mathcal{P}$, is a projector from the auxiliary quasiparticle local Hilbert space to the physical one:
\begin{equation}
    \mathcal{P} = \sum_{\Gamma n}\, \lambda_{\Gamma n}\, \ket{\Gamma } \bra{n}\,,\label{P}
\end{equation}
where $\ket{\Gamma }$ and $\ket{n }$ span, respectively, the physical 
and auxiliary local Hilbert spaces. The conventional Gutzwiller wavefunction corresponds to the physical and auxiliary Hilbert spaces having the same dimension. The ghost-Gutzwiller extension enlarges the auxiliary Hilbert space by introducing additional orbitals, the eponymous ``ghosts''. This provides a more flexible variational Ansatz, capable of capturing correlation features beyond the low-energy regime, such as the Hubbard bands~\cite{Lanata2017,Tagliente2025,Giuli2025,Pasqua2026}. \\
Upon introducing a positive semi-definite matrix $\hat P_0$ with components $P_{0\,nm} \equiv \langle \psi_\mathrm{qp} | m \rangle \langle n |\psi_\mathrm{qp} \rangle$, we can write the matrix $\hat{\lambda}$ with elements $\lambda_{\Gamma n}$ in \eqn{P} as $\hat \lambda = \hat\Phi\, \sqrt{\hat P_0^{-1}}$. In turn, the matrix elements of $\hat\Phi$ can be interpreted as the coefficients of a normalized impurity wavefunction
\begin{equation}
    \ket{\psi_\mathrm{imp}} = \sum_{\Gamma n}\, \Phi_{\Gamma n}\,
    \ket{\Gamma } \ket{\bar{n}}\,,
    \label{WFimpurity}
\end{equation}
where $\ket{\bar{n}}$ is a properly defined particle-hole transform of $\ket{n}$ \cite{Lanata2017}. A schematic representation of this impurity construction, and of the corresponding auxiliary quasiparticle bands, is shown in Fig.~\ref{fig:gGutWF}. The Hilbert space of the impurity  exactly matches the physical local Hilbert space, while the Hilbert space of the bath fermions coincides with the local Hilbert space of the auxiliary fermions that define $\ket{\psi_\text{qp}}$. \\
The variational character of the gGut wavefunction allows us to choose a restricted but physically motivated form of the impurity wavefunction, thereby simplifying the numerical optimization. Specifically, because of the constraint of no-double occupancy of the impurity, we just retain two bath sites, $a=1,2$, where 1 represents a bath pinned at the chemical potential, while bath 2 corresponds to an (almost) empty upper Hubbard band. The most general singlet impurity wavefunction, of which the two cases adopted below are limiting cases, is discussed in the SM, Sec.~\ref{sm:general-wf}. In this case, the GA implies that the operator $\mathcal{P}$ implements the following transmutation of the physical operators into auxiliary ones:
\be
\mathcal{P}^\dagger\, c^\dagga_{i\sigma}\mathcal{P} \to 
\sum_{a=1}^2\,R_a\, f^\dagga_{i a\sigma}\,,\quad
\mathcal{P}^\dagger\, \boldsymbol{S}_{i}\,\mathcal{P} \to 
\sum_{a\leq b}\,Q_{ab}\,\boldsymbol{S}_{iab}\,,
\label{mapping}
\ee
where $f^\dagga_{ia\sigma}$ are the annihilation operators of the auxiliary fermions, $\boldsymbol{S}_{iab}=\sum_{\alpha \beta}\, f^\dagger_{i a\alpha}\, \bd{S}_{\alpha \beta}\, f^\dagga_{i b\beta}$,   
while $R_a$ and $Q_{ab}=Q_{ba}^*$ are determined solely by the impurity wavefunction $\ket{\psi_\text{imp}}$ (see SM, Sec.~\ref{sm:mapping}). In addition, the GA entails the constraint $\bra{\psi_\text{imp}} f^\dagga_{b\sigma}\, f_{a\sigma}^\dagger \ket{\psi_\mathrm{imp}} \equiv \bra{\psi_\text{qp}} f_{ i a \sigma}^\dagger\, f^\dagga_{i b \sigma} \ket{\psi_\mathrm{qp}}$, with $f_{a\sigma}$ the bath operators. We note that this constraint implies that the almost empty upper Hubbard band of the impurity model corresponds to an almost full lower Hubbard band of the auxiliary lattice model, as schematically shown in Fig.~\ref{fig:gGutWF}. \\ 
This minimal two-bath parametrization of the impurity model suggests two distinct approaches to accomodating a finite hole density, as illustrated in Fig.~\ref{fig:gGutWF}.
\begin{itemize}
\item[A)]Similar to the conventional Gutzwiller solution, we can imagine the doped charge to be transferred solely to bath 1. This bath hybridizes with the impurity while the upper Hubbard band, bath 2, remains fully empty and decoupled. 
\item[B)] Alternatively, we can consider a scenario where the bath 1 remains strictly half-filled, unhybridized with the impurity but strongly spin-entangled with it. In this case the doped charge is transferred to bath 2 which is then promoted to the Fermi level. 
\end{itemize}
The case A) corresponds to the spin-singlet impurity wavefunction
\be
\ket{\psi_\text{imp}^\mathrm A} = \fract{\cos\theta}{\sqrt{2}}\,\big(
\ket{\up;\down,0} -\ket{\down;\up,0}\big)
+ \sin\theta\ket{0;2,0}\,,\label{psi-A}
\ee
where $\ket{\gamma_\text{imp};\gamma_1,\gamma_2}$, 
$\gamma=0,\up,\down,2$, are basis states, and $\cos^2\theta=1-\delta$ is fixed by the physical doping.
The wavefunction \eqn{psi-A} implies that $R_2=Q_{22}=Q_{12}=0$ (see SM, Sec.~\ref{sm:mapping}) and
\beal
 R_1&= \sqrt{ \fract{2 \delta}{\;1 + \delta\;}\; } \,,& Q_{11} &= \fract{2}{\;1+\delta\;}\;. \label{RQ-A}
\eal
The uncorrelated Fock state $\ket{\psi_\text{qp}}$ must be chosen so as to minimize the expectation value of the quasiparticle Hamiltonian 
\beal
H_\text{qp}^\mathrm{A} &= -R_1^2\,\sum_{ij\sigma}\,t_{ij}\,\big(
f^\dagger_{i1\sigma}\,f^\dagga_{j1\sigma} + \text{H.c.}\big)\\
&\quad -\mu\,\sum_{i\sigma}\,f^\dagger_{i1\sigma}\,f^\dagga_{i1\sigma}
 + J\,Q_{11}^2\,\sum_{\langle ij \rangle}\,\boldsymbol{S}_{i1} \cdot \boldsymbol{S}_{j1}\,,\label{Ham-A}
\eal
where $\boldsymbol{S}_{i1}=\boldsymbol{S}_{i11}$ and $\mu$ is a Lagrange multiplier that enforces $\langle\,\sum_\sigma\,f^\dagger_{i1\sigma}\,f^\dagga_{i1\sigma}\,\rangle=1-\delta$. We observe that \eqn{Ham-A} is equivalent to the conventional Gutzwiller approach to the $t$-$J$ model \cite{ZhangGrosRiceShiba1988}. In fact, the results we are going to show recover those in \cite{ZhangGrosRiceShiba1988,LiuKotliar1988}.
In particular, at low doping $\ket{\psi_\text{qp}}$ is a BCS wavefunction with $d$-wave symmetry \cite{ZhangGrosRiceShiba1988,LiuKotliar1988}, 
$\langle f^\dagger_{1\bk\up}\,f^\dagger_{1-\bk\down}\rangle = \Delta_\text{SC}\,(
\cos k_x - \cos k_y)$, which we denote as SC state. The order parameter 
$\Delta_\text{SC}$, assumed real and positive, is maximum at $\delta=0$ and decreases monotonically with increasing $\delta$. The 
physical order parameter, $\Delta=R_1^2\,\Delta_\text{SC}$, is instead non-monotonic and vanishes at $\delta=0$. Above a critical doping, 
$\Delta_\text{SC}=0$ \cite{ZhangGrosRiceShiba1988,LiuKotliar1988} and 
$\ket{\psi_\text{qp}}$ describes a normal Fermi liquid, which we denote as FL, with a large electron-like Fermi surface that accommodates $1-\delta$ electrons per site.
\\
\noindent
On the contrary, case B) corresponds to the spin-singlet impurity wavefunction
\beal
\ket{\psi_\text{imp}^\mathrm{B}} &= \fract{\cos\theta}{\sqrt{2}}\,\big(
\ket{\up;\down,0} -\ket{\down;\up,0}\big)\\
&\qquad + \fract{\sin\theta}{\sqrt{2}}\,\big(
\ket{0;\up,\down} -\ket{0;\down,\up}\big)\,,\label{psi-B}
\eal
where, as before, $\cos^2\theta=1-\delta$, while now $R_1=Q_{22}=Q_{12}=0$ and 
\beal
R_2 &=\sqrt{\fract{1-\delta}{\;2-\delta\;}\;}\;,&
Q_{11} &= 2\,(1-\delta)\,.\label{RQ-B}
\eal
The quasiparticle Hamiltonian in this case reads 
\beal
H_\text{qp}^\mathrm{B} &= -R_2^2\,\sum_{ij\sigma}\,t_{ij}\,\big(
f^\dagger_{i2\sigma}\,f^\dagga_{j2\sigma} + \text{H.c.}\big)\\
&\quad -\mu\,\sum_{i\sigma}\,f^\dagger_{i2\sigma}\,f^\dagga_{i2\sigma}
 + J\,Q_{11}^2\,\sum_{\langle ij \rangle}\,\boldsymbol{S}_{i1} \cdot \boldsymbol{S}_{j1}\\
 &= H_2^\mathrm{B}+ H_1^\mathrm{B}
 \,,\label{Ham-B}
\eal
where $\mu$ enforces $\langle\,\sum_\sigma\,f^\dagger_{i2\sigma}\,f^\dagga_{i2\sigma}\,\rangle=2-\delta$, while the orbital 1 is half-filled. Consequently, the uncorrelated wavefunction factorizes as $\ket{\psi_\text{qp}} = \ket{\psi_1}\ket{\psi_2}$ with both $\ket{\psi_1}$ and $\ket{\psi_2}$ Fock states that minimize the expectation value of 
$H_1^\mathrm{B}$ and $H_2^\mathrm{B}$, respectively, 
with site-independent local single-particle density matrices. This implies that $\ket{\psi_1}$ is not simply the ground state of the Heisenberg model $H_1^\mathrm{B}$, as this is not a Fock state.
In reality, the wavefunction $\ket{\psi_1}$ that minimizes the Heisenberg exchange and is consistent with the GA describes a $\pi$-flux state \cite{Affleck-Marston-flux-PRB1988} or any other equivalent state under the $SU(2)$ gauge symmetry possessed by $H_1^\mathrm{B}$ \cite{Affleck-flux-PRB1988,LiuKotliar1988}, such as a $d$-wave BCS state. For simplicity, we henceforth refer to this state as a $\pi$-flux one with order parameter $\Delta_{\pi\text{-flux}}\in \mathbb{R}^+$ that coincides with the above $\Delta_\text{SC}$ at $\delta=0$. 
We emphasize that such gauge symmetry simply reflects an invariance property of the 
Gutzwiller wavefunction \eqn{G-wavefunction}, $\mathcal{P}_i\to \mathcal{P}_i\,U^\dagger_i$ and $\ket{\psi_\text{qp}}\to \prod_i U_i^\dagga \ket{\psi_\text{qp}}$, with unitary $U_i$, which becomes manifest when the auxiliary fermions have $R=0$. The wavefunction $\ket{\psi_2}$ corresponds instead to a conventional metal with a hole-like Fermi surface that accommodates $2-\delta$ electrons per site. In other words, $\ket{\psi_\text{qp}} = \ket{\psi_1}\ket{\psi_2}$ describes precisely a $\mathrm{FL}^{\!*}$ state, with a band of conventional quasiparticles with Fermi surface violating Luttinger's theorem and a band of quasiparticles without spectral weight in the physical electron, $R_1=0$, which represent the spinons. \\
\noindent
The variational approach we have so far described can be further extended at finite temperature $T\neq 0$ by employing the well-known result that the free energy $F$ associated with the Hamiltonian $H$ satisfies 
\beal
F &= \min_\rho\Big( U(\rho) -T\,S(\rho)\Big)\\
&= \min_\rho\Big( \Tr\big(\rho\,H\big) +T\,\Tr\big(\rho\,\ln\rho\big)\Big)\,,
\label{F:def}
\eal
where $\rho$ spans the space of all density matrices.  
Restricting the search to a subspace of all possible $\rho$ thus provides an upper bound of the true free-energy. Here, we assume a density matrix \cite{Sandri2013}
\begin{equation}
    \rho_\text{gGut} = \prod_{i}\, \mathcal{P}_i\, \rho_\mathrm{qp}\, \mathcal{P}_i^\dagger\,,
    \label{density-matrix}
\end{equation}
where $\rho_\mathrm{qp}$ is the Boltzmann weight associated with the 
quasiparticle Hamiltonians \eqn{Ham-A} or \eqn{Ham-B}. Since a closed expression for $S(\rho)$ cannot be found within the GA, we use a lower bound that extends the one in \cite{Sandri2013} and becomes exact in both $T=0$ and $T\to\infty$ limits (see SM, Sec.~\ref{sm:entropy} for the derivation):
\begin{equation}
    S(\rho_\text{gGut}) \geq S(\rho_\mathrm{qp}) \left( 1 - \frac{1}{\; N \ln{d}\;}\; \Tr{\left( \rho_\mathrm{qp} \mathcal{Q}\, \ln{\mathcal{Q}} \right)} \right)\,,
    \label{lower bound}
\end{equation}
with $S(\rho_\mathrm{qp})$ the entropy of the quasiparticle density matrix, $\mathcal{Q} = \prod_{i}\, \mathcal{P}_i^\dagger\, \mathcal{P}_i$, $N$ the number of sites, and $d$ the dimension of the local auxiliary Hilbert space. The variational optimization thus reduces to finding the minimum 
of \eqn{F:def} within the subspace spanned by $\rho_\text{gGut}$ in \eqn{density-matrix} with the entropy replaced by the lower bound \eqn{lower bound}. The minimum so obtained provides an upper bound of the actual free energy. We emphasize that the impurity wavefunctions \eqn{psi-A} and \eqn{psi-B} are free from any parameters. Consequently, the variational optimization simplifies to the Hartree-Fock decoupling of the exchange terms in \eqn{Ham-A} and \eqn{Ham-B} as previously anticipated.\\
We also stress that the present variational approach gives only access to static thermodynamic quantities but not to dynamical ones. Among the former is the uniform static spin susceptibility $\chi$, discussed below, which can be directly compared with the NMR Knight shift of cuprates~\cite{MilaRice1989,Alloul1989,Takigawa1991}. By contrast, the NMR relaxation rate $1/T_1$ \cite{Moriya1963,MillisMonienPines1990} as well as transport and spectral properties lie beyond the reach of the present framework and are left for future work. Specifically, to compute $\chi$ we add to the Hamiltonian \eqn{Ham} a Zeeman term $-h\,\sum_i\,S^z_i$ and generalize the impurity wavefunctions \eqn{psi-A} and \eqn{psi-B} to allow for a finite uniform magnetization $m_z = \langle S^z_i\rangle$ (see SM, Sec.~\ref{sm:magnetization}). The susceptibility then follows from $\chi = \partial m_z/\partial h\,|_{h=0}$.\\
\begin{figure}[t]
    \centering
    \includegraphics[width=0.5\textwidth]{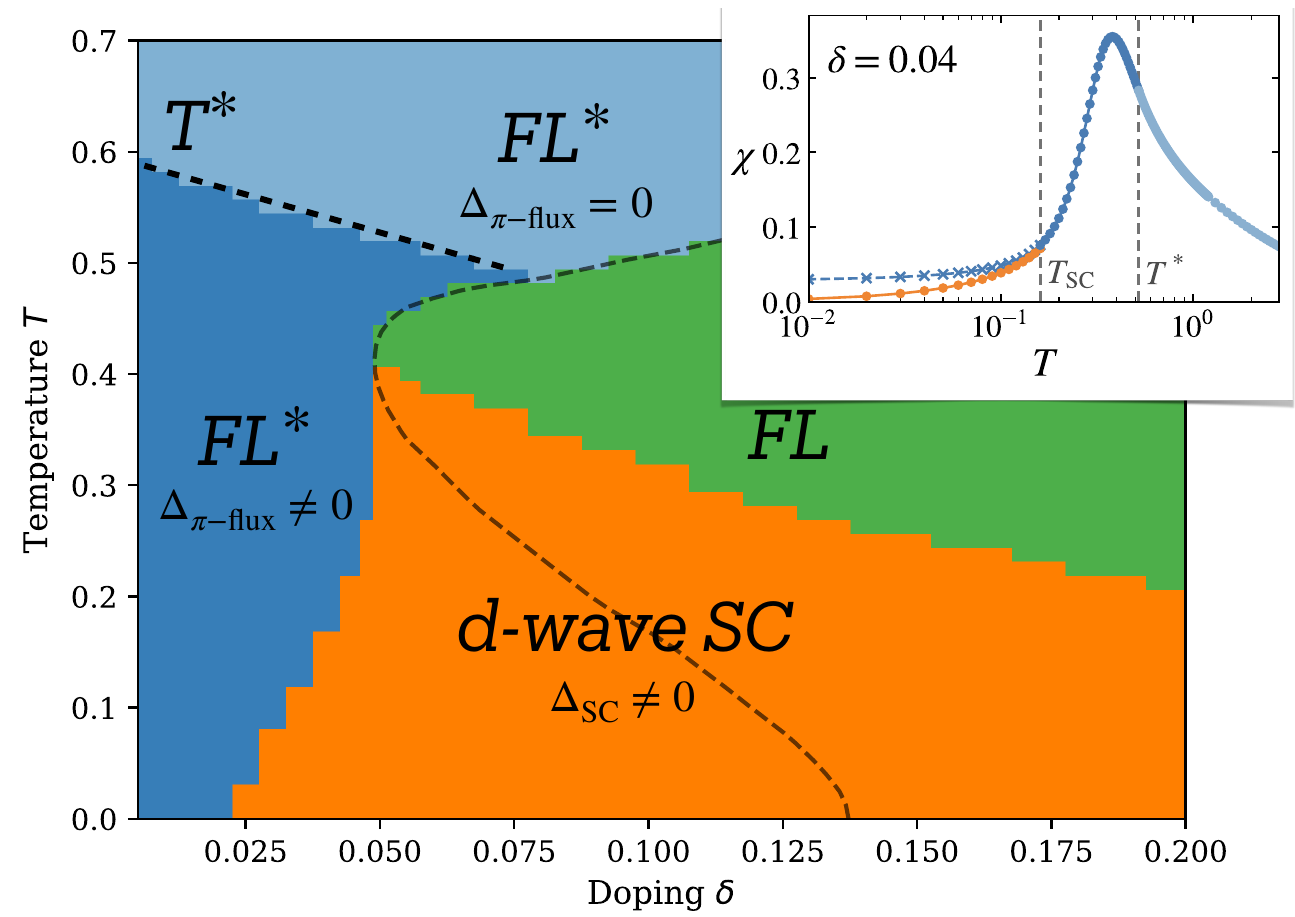}
    \caption{Temperature--doping phase diagram of the $t$-$J$ model
    ($J/t=0.4$, $t^\prime/t=-0.4$, $t=1$) within the ghost-Gutzwiller Ansatz.
    At low doping the normal state is a fractionalized Fermi liquid
    ($\mathrm{FL}^{\!*}$) with a small, hole-like Fermi surface accounting just for the doped holes; the overdoped normal
    state is a conventional FL with a large, electron-like Fermi surface satisfying Luttinger's count. The gray dashed line marks the boundary between the two metallic states preventing superconductivity. Within the $\mathrm{FL}^{\!*}$ region, the black dashed line is the spinon pairing scale $T^{\!*}$, below which the $\pi$-flux order parameter $\Delta_{\pi\text{-flux}}$
    is finite. When superconductivity is allowed, a $d$-wave superconducting dome ($\Delta_{\mathrm{SC}}\neq0$) intrudes between the two metallic regimes. Inset: uniform static spin susceptibility $\chi$ as a function of $T$ at $\delta=0.04$. Below $T_\mathrm{SC}$, the solid line is the superconducting solution and the dashed line the $\pi$-flux $\mathrm{FL}^{\!*}$ obtained when superconductivity is prevented.}
    \label{fig:PhaseDiagram}
\end{figure}
\textit{Phase diagram.}---We are now ready to discuss the variational phase diagram. The parameters in \eqn{Ham} are chosen to  represent the one-band description of  cuprates \cite{Pavarini2001,Shih2004}: Heisenberg coupling $J/t=0.4$, nearest-neighbor hopping $t=1$, and next-nearest-neighbor hopping $t^\prime/t=-0.4$. Varying $J$ and $t^\prime$ shifts the transition lines, but leaves the phase diagram qualitatively unchanged (see SM, Sec.~\ref{sm:parameters}).
The resulting temperature-doping phase diagram is shown in Fig.~\ref{fig:PhaseDiagram}. At low doping, the optimal impurity wavefunction is of the type B \eqn{psi-B} and yields a $\text{FL}^*$ phase. The $\pi$-flux order parameter 
$\Delta_{\pi\text{-flux}}$ is finite below $T^{\,\!*}\propto Q_{11}^2\, J \sim (1-\delta)^2$ and zero above. If superconductivity is prevented, the $\text{FL}^*$ transitions above a critical doping into a conventional FL state characterized by the A-type impurity wavefunction \eqn{psi-A}.  This is shown by the dashed line in Fig.~\ref{fig:PhaseDiagram} and corresponds to a reconstruction of the Fermi surface from hole-like to electron-like.  Allowing for superconductivity results in the SC dome depicted in Fig.~\ref{fig:PhaseDiagram}, which is again associated with the impurity wavefunction \eqn{psi-A}. As in Refs.~\cite{LiuKotliar1988,ZhangGrosRiceShiba1988}, we regard the finite temperature SC phase as physically meaningful: although forbidden in a strictly two dimensional model, superconducting long-range order is stabilized in cuprates by the weak coupling between CuO$_2$ planes, with the actual $T_c$ well approximated by its mean-field value calculated in two dimensions. In the slave-boson mean-field theory of the $t$-$J$ model \cite{LiuKotliar1988}, the superconducting dome emerges from the interplay of two distinct temperatures: the slave-boson condensation temperature, which is zero at $\delta=0$ and rises with doping, and the temperature at which the quasiparticle $d$-wave order parameter $\Delta_\text{SC}$ vanishes, which decreases monotonically with increasing $\delta$. In our calculation, the dome is a result of the competition between the larger internal energy of the SC phase and the larger entropy of the $\text{FL}^*$ and FL states, since both of which have Fermi surfaces.\\
The transition lines between the SC/FL and the $\text{FL}^*$ phases arise from the two distinct impurity wavefunctions \eqn{psi-A} and \eqn{psi-B}, respectively. Consequently, these transitions mark the crossing of their respective free energies and are therefore first-order. We anticipate that optimizing the variational free-energy with the most general impurity wavefunction incorporating \eqn{psi-A} and \eqn{psi-B} as special cases would modify the transition lines and the precise shape of the phase diagram, but not its overall features. In particular, it could change the order of the transitions, as well as the way the different phases merge into one another. This more involved calculation, although valuable, exceeds the scope of this work which is to demonstrate the potential of the ghost-Gutzwiller approximation to capture exotic phases such as $\text{FL}^*$ at extremely low computational cost. \\ 
We conclude by discussing the uniform static spin susceptibility $\chi$, shown in the inset of Fig.~\ref{fig:PhaseDiagram} as a function of $T$ for $\delta=0.04$. In the flux-less $\mathrm{FL}^{\!*}$, the spinons have a large Fermi surface. Below $T^{\,\!*}$, the $\pi$-flux order parameter shrinks it to nodal points, thereby depleting the spinon density of states. The susceptibility is  the sum of the Pauli-like contribution of the hole pockets, which keeps growing upon cooling until it saturates, and the spinon one, which is instead suppressed. As a result, $\chi$ reaches a maximum below $T^{\,\!*}$ and then drops markedly, matching
well the drop of the NMR Knight shift observed in underdoped cuprates.
Below $T_\mathrm{SC}$, $d$-wave pairing suppresses $\chi$ further and makes it vanish as $T\to0$. On the contrary, if superconductivity is prevented, $\chi$ saturates to a small but finite value set by the hole pockets, as recently observed in underdoped YBa$_2$Cu$_3$O$_y$ at high magnetic fields~\cite{Zhou2025}. This rich phenomenology, which we can capture within a very simple Ansatz and is consistent with experimental evidence, can be reproduced 
by cluster and diagrammatic extensions of DMFT
at much higher computational cost \cite{Chen2017,Sordi2012,Macridin2006,Malcolms2026}. \\
\textit{Conclusions.}---We have demonstrated that the ghost-Gutzwiller approximation, applied to the single-band $t$-$J$ model, provides a straightforward variational route to a fractionalized Fermi liquid ($\mathrm{FL}^{\!*}$). This phase is characterized by neutral yet spinful quasiparticles coexisting with conventional ones, which form hole-like Fermi pockets whose volume is determined solely by the doping away from half-filling.
\\
Despite its minimal structure, the Ansatz successfully reproduces the essential qualitative features of the cuprate phase diagram: a $\mathrm{FL}^{\!*}$ metal, a $d$-wave superconducting dome, and an overdoped conventional Fermi liquid with a large Fermi surface. In the $\mathrm{FL}^{\!*}$ regime, the spinons arrange into a $\pi$-flux phase, which is gauge equivalent to a $d$-wave condensate, below a temperature scale $T^{\,\!*}$ associated with the onset of pseudogap behavior. This does not, however, directly produce a physical superconducting order parameter. Superconductivity emerges only when the paired spinons acquire physical electron spectral weight through the Gutzwiller projection, that is, upon their recombination into physical electrons.
\\
We emphasize that stabilizing the $\mathrm{FL}^{\!*}$ phase relies on the
nonlocal character of the spin-exchange interaction, which endows the
self-energy with a genuine momentum dependence that a purely local interaction
would not produce within the Gutzwiller approximation. This is, however, not equivalent to a conventional mean-field decoupling of the exchange term: in the present case, the interaction is first dressed by the Gutzwiller projector $\mathcal{P}$, which renormalizes the effective quasiparticle dynamics and may also induce a frequency dependent self-energy. We mention that preliminary calculations suggest that an analogous $\mathrm{FL}^{\!*}$ solution can persist as a metastable state also in the doped single-band Hubbard model deep in the Mott regime, which we leave for future work.\\
\noindent
In a broader context, the flexibility and low cost of the ghost-Gutzwiller Ansatz render it a promising tool for investigating $\mathrm{FL}^{\!*}$ and related unexplored phases in multi-orbital models of correlated materials. This approach may offer a complementary variational perspective to more computationally expensive methods, such as cluster extensions of dynamical mean-field theory~\cite{Hettler1998,Kotliar2001,Maier2005} and variational Monte Carlo~\cite{Gros1988,BeccaSorella2017,Rende2026}, which can study similar fractionalized regimes but at significantly higher computational expense. This comparison may aid in clarifying, for instance, the microscopic origin of experimentally observed insulators with gapless neutral excitations despite the finite charge gap~\cite{Dzero2016,Tan2015,Hartstein2018}.

\textit{Acknowledgments.}---
We acknowledge insightful discussions with S. Giuli, D. Poilblanc, G. Sangiovanni, and A. M. Tagliente.

%

\onecolumngrid
\clearpage

\setcounter{equation}{0}
\setcounter{figure}{0}
\setcounter{table}{0}
\setcounter{section}{0}
\renewcommand{\theequation}{S\arabic{equation}}
\renewcommand{\thefigure}{S\arabic{figure}}
\renewcommand{\thetable}{S\arabic{table}}

\setcounter{secnumdepth}{3}
\renewcommand{\thesection}{\Roman{section}}

\begin{center}
    \large\textbf{Supplemental Material to:\\ "Fractionalized Fermi liquids with the ghost-Gutzwiller Ansatz"}
\end{center}

\vspace{1em}

\begingroup
\setcounter{tocdepth}{3}
\makeatletter
\let\oldcontentsline\contentsline
\renewcommand{\contentsline}[4]{%
    \def\tmp@text{#2}%
    \def\tmp@ref{\numberline {}References}%
    \ifnum\pdfstrcmp{#1}{title}=0\relax\else
      \ifnum\pdfstrcmp{#1}{abstract}=0\relax\else
        \ifx\tmp@text\tmp@ref\relax\else
          \oldcontentsline{#1}{#2}{#3}{#4}%
        \fi
      \fi
    \fi}
\makeatother
\tableofcontents
\endgroup

\vspace{1em}

\section{General spin-singlet impurity wavefunction}
\label{sm:general-wf}
\noindent
The wavefunctions $\ket{\psi_\text{imp}^\mathrm A}$ and $\ket{\psi_\text{imp}^\mathrm B}$ discussed in the main text can be derived as two limiting cases of the most general spin-singlet impurity wavefunction with two bath orbitals. Keeping the notation $\ket{\gamma_\mathrm{imp};\gamma_1,\gamma_2}$ with $\gamma=0,\up,\down,2$, and exploiting the no-double-occupancy constraint on the physical site together with $SU(2)$ symmetry, the general spin-singlet wavefunction reads
\begin{equation}
    \begin{aligned}
    \ket{\psi_\mathrm{imp}}
    &= \frac{\alpha_1}{\sqrt2}\big(\ket{\up;\down,0}-\ket{\down;\up,0}\big)
    + \frac{\alpha_2}{\sqrt2}\big(\ket{\up;0,\down}-\ket{\down;0,\up}\big)\\
    &\quad + \frac{\alpha_3}{\sqrt2}\big(\ket{0;\up,\down}-\ket{0;\down,\up}\big)
    + \alpha_4\ket{0;2,0} + \alpha_5\ket{0;0,2}\,,
    \end{aligned}
    \label{psi-imp-full}
\end{equation}
where the amplitudes $\alpha_i$ can be taken real. These amplitudes are subject to three constraints: the wavefunction normalization 
\begin{equation}
    \sum_i\alpha_i^2=1\,,
\end{equation}
the physical electron density 
\begin{equation}
    \alpha_1^2+\alpha_2^2=1-\delta\,,
\end{equation}
and the requirement, without loss of variational freedom, that the bath orbitals are in their natural basis,
\begin{equation}
    \bra{\psi_\mathrm{imp}} f^\dagger_{1\sigma}f^\dagga_{2\sigma}\ket{\psi_\mathrm{imp}}
    = \frac{\alpha_1\alpha_2}{2} + \frac{\alpha_3(\alpha_4+\alpha_5)}{\sqrt2} = 0\,.
    \label{natural-basis}
\end{equation}
The impurity wavefunction \eqn{psi-imp-full} therefore carries only two independent variational parameters, on top of the choice of $\ket{\psi_\mathrm{qp}}$. \\
\noindent
The two cases discussed in the main text are recovered as special points of this family: case A corresponds to 
\begin{equation}
    \alpha_1=\sqrt{1-\delta}\,, \quad \alpha_4=\sqrt{\delta}\,, \quad \alpha_2=\alpha_3=\alpha_5=0\,,
\end{equation}
and case B instead corresponds to 
\begin{equation}
    \alpha_1=\sqrt{1-\delta}\,, \quad \alpha_3=\sqrt{\delta}\,, \quad \alpha_2=\alpha_4=\alpha_5=0\,.
\end{equation}

\section{Operators mapping and renormalization factors}
\label{sm:mapping}
\noindent
The transmutation of the physical operators into auxiliary ones discussed in Eq.~\eqn{mapping} is not an operator identity. Rather, it has to be understood as valid inside the expectation value of the physical Hamiltonian \eqn{Ham} over the ghost-Gutzwiller wavefunction \eqn{G-wavefunction}. Within the Gutzwiller approximation, which becomes exact in the limit of infinite lattice coordination, this expectation value is obtained by replacing each physical operator with its renormalized auxiliary counterpart and evaluating the result on the uncorrelated Fock state $\ket{\psi_\mathrm{qp}}$,
\begin{equation}
    \frac{\bra{\Psi_\mathrm{gGut}}\, H\, \ket{\Psi_\mathrm{gGut}}}{\langle \Psi_\mathrm{gGut} | \Psi_\mathrm{gGut} \rangle}
    = \frac{\bra{\psi_\mathrm{qp}}\, H_\mathrm{qp}\, \ket{\psi_\mathrm{qp}}}{\langle \psi_\mathrm{qp} | \psi_\mathrm{qp} \rangle}\,,
    \label{exp-value}
\end{equation}
where $H_\mathrm{qp}$ is the quasiparticle Hamiltonian obtained from $H$ through the mapping \eqn{mapping}, see Eqs.~\eqn{Ham-A} and \eqn{Ham-B}. The equality \eqn{exp-value} holds as the physical Hamiltonian \eqn{Ham} contains no local interaction terms. The renormalization factors $R_a$ and $Q_{ab}$ in Eqs.~\eqn{Ham-A} and \eqn{Ham-B} are fixed by the impurity wavefunction $\ket{\psi_\mathrm{imp}}$, so that the variational problem is entirely mapped onto finding the optimal uncorrelated state $\ket{\psi_\mathrm{qp}}$ that minimizes the expectation value of the quasiparticle Hamiltonian, subject to the GA constraint relating the local density matrices of $\ket{\psi_\mathrm{imp}}$ and $\ket{\psi_\mathrm{qp}}$. \\
We now show explicitly how the renormalization factors $R_a$ and $Q_{ab}$ entering the mapping \eqn{mapping} are determined from the impurity wavefunction.
We start from case A \eqn{psi-A}, where using $\hat\lambda=\hat\Phi\,\sqrt{\hat P_0^{-1}}$ and the natural basis where the uncorrelated local density matrix is diagonal, 
\begin{equation}
    \lambda_{\Gamma m}=\Phi_{\Gamma m} \,\sqrt{ P_0^{-1}(m)}\,,
\end{equation}
we readily obtain the explicit form of $\mathcal{P}_\mathrm{A}$ as
\begin{equation}
    \mathcal{P}_\mathrm{A} = \sqrt{\frac{1-\delta}{2 \, P_0(\sigma, 2)}} \sum_\sigma \ket{\sigma}\bra{\sigma, 2} + \sqrt{\frac{\delta}{P_0(0, 2)}} \, \ket{0}\bra{0, 2}\,,
\end{equation}
where $m \equiv (\gamma_1,\gamma_2)$ labels the local configuration of the two auxiliary orbitals in $\ket{\psi_\text{qp}}$, with probability $P_0(\gamma_1,\gamma_2) = |\bra{\gamma_1,\gamma_2}\, \psi_\text{qp}\rangle|^2$ . We stress that these auxiliary configurations are the particle-hole transform of the bath states appearing in the impurity wavefunctions \eqn{psi-A} and \eqn{psi-B}; in particular, the upper Hubbard band orbital 2, empty in $\ket{\psi_\text{imp}}$, is full in $\ket{\psi_\text{qp}}$, e.g.\ $n^\mathrm{qp}_{2\sigma}=1$ in case A. We note that the two terms $\ket{\sigma}\bra{\sigma,2}$ enter with the same sign for both spin components, as required by the $SU(2)$ invariance of $\mathcal{P}_\mathrm{A}$.
From $\ket{\psi_\mathrm{imp}^\mathrm{A}}$ in Eq.~\eqn{psi-A} and the GA constraint 
\begin{equation}
    \bra{\psi_\text{imp}} f^\dagga_{b\sigma}\, f_{a\sigma}^\dagger \ket{\psi_\mathrm{imp}} \equiv \bra{\psi_\text{qp}} f_{ i a \sigma}^\dagger\, f^\dagga_{i b \sigma} \ket{\psi_\mathrm{qp}} \, ,
\end{equation}
it follows that the auxiliary orbital 1 has an average occupation number per spin $n^\mathrm{qp}_{1 \sigma}=(1-\delta)/2$, while orbital 2 has $n^\mathrm{qp}_{2 \sigma}=1$; hence
\begin{equation}
    \begin{aligned}
        P_0(\sigma,2)&=n^\mathrm{qp}_{1 \sigma} \, \left( 1 - n^\mathrm{qp}_{1 \bar \sigma} \right) \, n^\mathrm{qp}_{2 \sigma} \, n^\mathrm{qp}_{2 \bar \sigma}=\frac{1-\delta^2}{4}\,,\\
        P_0(0,2)&=(1 - n^\mathrm{qp}_{1 \sigma}) \, \left( 1 - n^\mathrm{qp}_{1 \bar \sigma} \right) \, n^\mathrm{qp}_{2 \sigma} \, n^\mathrm{qp}_{2 \bar \sigma}=\frac{(1+\delta)^2}{4}\,,
    \end{aligned}
\end{equation}
with $\bar\sigma$ the spin opposite to $\sigma$. \\
\noindent
Applying these projectors to the physical creation operator $c^\dagger_{\sigma} = \ket{\sigma}\bra{0} + \ket{2}\bra{\bar \sigma}$ selects the only non-vanishing term,
\begin{equation}
    \mathcal{P}_\mathrm{A}^{\dagger}\, c^\dagger_{\sigma}\,\mathcal{P}_\mathrm{A}
    = \sqrt{\frac{(1-\delta)\,\delta}{2 \, P_0(\sigma, 2)\, P_0(0, 2)}}\; \ket{\sigma, 2}\bra{0, 2}\,.
\end{equation}
The operator acting on the local auxiliary Hilbert space can be written exactly as
\begin{equation}
    \ket{\sigma, 2}\bra{0, 2}=f^\dagger_{1\sigma}\,(1- n_{1\bar\sigma})\, n_{2\up} n_{2\down}.
\end{equation}
Within the GA the local density operators are replaced by their expectation values on $\ket{\psi_\text{qp}}$, so that
\begin{equation}
    \ket{\sigma, 2}\bra{0, 2} \;\to\; (1-n^\mathrm{qp}_{1\bar\sigma})\,f^\dagger_{1\sigma}\,,
\end{equation}
and therefore
\begin{equation}
    \mathcal{P}_\mathrm{A}^{\dagger}\, c^\dagger_{\sigma}\,\mathcal{P}_\mathrm{A}
    \to \sqrt{\frac{2\delta}{1+\delta}} \, f^\dagger_{1\sigma} \equiv R_1\, f^\dagger_{1\sigma} \,,
\end{equation}
in agreement with Eq.~\eqn{RQ-A}.\\
The spin renormalization factor is obtained analogously: Projecting the physical raising operator $S^+=\ket{\up}\bra{\down}$, only the term connecting the two singlet components of Eq.~\eqn{psi-A} survives,
\begin{equation}
    \mathcal{P}_\mathrm{A}^{\dagger}\, S^+\,\mathcal{P}_\mathrm{A}
    = \frac{1-\delta}{2 \, P_0(\sigma, 2)}\; \ket{\up, 2}\bra{\down, 2}\,.
\end{equation}
Recalling
\begin{equation}
    \ket{\up, 2}\bra{\down, 2}=f^\dagger_{1\up}f^\dagga_{1\down}\, n_{2\up} n_{2\down}=S^+_{i11}\, n_{2\up} n_{2\down}
\end{equation}
and replacing the orbital-2 density operators by their expectation value $n^\mathrm{qp}_{2\up} n^\mathrm{qp}_{2\down}=1$ gives
\begin{equation}
    \mathcal{P}_\mathrm{A}^{\dagger}\, S^+\,\mathcal{P}_\mathrm{A}
    \to \frac{2}{1+\delta} \, S^+_{i11} \equiv Q_{11}\, S^+_{i11}\,.
\end{equation}
By $SU(2)$ symmetry the same factor $Q_{11}$ renormalizes every component of the spin operator $\boldsymbol{S}_{i11}$, in agreement with Eq.~\eqn{RQ-A}.\\
The same analysis applies to case B \eqn{psi-B}: In this case the local linear operator $\mathcal{P}_\mathrm{B}$ reads
\begin{equation}
    \mathcal{P}_\mathrm{B} = \sqrt{\frac{1-\delta}{2 \, P_0(\sigma, 2)}} \sum_\sigma \ket{\sigma}\bra{\sigma, 2} + \sqrt{\frac{\delta}{2 \, P_0(\sigma, \bar\sigma)}} \, \big(\ket{0}\bra{\up, \down}-\ket{0}\bra{\down,\up}\big)\,,
\end{equation}
where now the empty impurity state is paired with the spin singlet of the two singly-occupied auxiliary orbitals, again an $SU(2)$ scalar. Now the auxiliary orbital 1 is half-filled, $n^\mathrm{qp}_{1\sigma}=1/2$, while orbital 2 hosts $2-\delta$ electrons, $n^\mathrm{qp}_{2\sigma}=(2-\delta)/2$; hence
\begin{equation}
    \begin{aligned}
        P_0(\sigma,2)&=n^\mathrm{qp}_{1 \sigma} \left( 1 - n^\mathrm{qp}_{1 \bar \sigma} \right) n^\mathrm{qp}_{2 \sigma} \, n^\mathrm{qp}_{2 \bar \sigma}=\frac{(2-\delta)^2}{16}\,,\\
        P_0(\sigma,\bar\sigma)&=n^\mathrm{qp}_{1 \sigma} \left( 1 - n^\mathrm{qp}_{1 \bar \sigma} \right) n^\mathrm{qp}_{2 \bar\sigma} \left( 1 - n^\mathrm{qp}_{2 \sigma} \right)=\frac{(2-\delta)\delta}{16}\,.
    \end{aligned}
\end{equation}
Acting on the physical creation operator now gives
\begin{equation}
    \mathcal{P}_\mathrm{B}^{\dagger}\, c^\dagger_{\sigma}\,\mathcal{P}_\mathrm{B}
    = \sqrt{\frac{(1-\delta)\,\delta}{4 \, P_0(\sigma, 2)\, P_0(\sigma,\bar\sigma)}}\; \ket{\sigma, 2}\big(\bra{\up, \down}-\bra{\down,\up}\big)\,.
\end{equation}
This operator enters the inter-site hopping term of $H_\mathrm{qp}^\mathrm{B}$ \eqn{Ham-B}, paired with its Hermitian conjugate on a neighboring site. Of its two contributions, only 
\begin{equation}
    \ket{\sigma, 2}\bra{\sigma,\bar\sigma}=f^\dagger_{2\sigma}\, n_{1\sigma}(1-n_{1\bar\sigma})\, n_{2\bar\sigma}
\end{equation}
creates a single auxiliary fermion in orbital 2 and contributes to the renormalized hopping. The latter, 
\begin{equation}
    \ket{\sigma, 2}\bra{\bar\sigma,\sigma}=f^\dagger_{2\bar\sigma}\, n_{2\sigma}\,f^\dagger_{1\sigma}f^\dagga_{1\bar\sigma} \, ,
\end{equation}
additionally contains the on-site spin flip $f^\dagger_{1\sigma}f^\dagga_{1\bar\sigma}$ of orbital 1; within the GA this local factor multiplies the inter-site contraction and is replaced by its expectation value on the paramagnetic state $\ket{\psi_\text{qp}}$, $\langle f^\dagger_{1\sigma} f^\dagga_{1\bar\sigma}\rangle_\mathrm{qp}=0$, so that this term does not contribute to the energy. Furthermore, the auxiliary orbitals are also natural orbitals, then the off-diagonal contraction $\langle f^\dagger_{2\bar\sigma} f^\dagga_{1\bar\sigma}\rangle_\mathrm{qp}$ vanishes as well. Replacing the remaining densities by their expectation values gives
\begin{equation}
    \ket{\sigma, 2}\bra{\sigma,\bar\sigma} \;\to\; n^\mathrm{qp}_{1\sigma}(1-n^\mathrm{qp}_{1\bar\sigma})\, n^\mathrm{qp}_{2\bar\sigma}\, f^\dagger_{2\sigma}\,,
\end{equation}
and therefore
\begin{equation}
    \mathcal{P}_\mathrm{B}^{\dagger}\, c^\dagger_{\sigma}\,\mathcal{P}_\mathrm{B}
    \to \sqrt{\frac{1-\delta}{2-\delta}} \, f^\dagger_{2\sigma} \equiv R_2\, f^\dagger_{2\sigma}\,,
\end{equation}
in agreement with Eq.~\eqn{RQ-B}. \\
The spin factor follows as in case A,
\begin{equation}
    \mathcal{P}_\mathrm{B}^{\dagger}\, S^+\,\mathcal{P}_\mathrm{B}
    = \frac{1-\delta}{2 \, P_0(\sigma, 2)}\; \ket{\up, 2}\bra{\down, 2}\,.
\end{equation}
With 
\begin{equation}
    \ket{\up, 2}\bra{\down, 2}=f^\dagger_{1\up}f^\dagga_{1\down}\, n_{2\up} n_{2\down}=S^+_{i11}\, n_{2\up} n_{2\down}
\end{equation} 
and replacing the orbital-2 density operators by $n^\mathrm{qp}_{2\up} n^\mathrm{qp}_{2\down}=((2-\delta)/2)^2$, we obtain
\begin{equation}
    \mathcal{P}_\mathrm{B}^{\dagger}\, S^+\,\mathcal{P}_\mathrm{B}
    \to 2(1-\delta) \, S^+_{i11} \equiv Q_{11}\, S^+_{i11}\,.
\end{equation}
By $SU(2)$ symmetry the same factor $Q_{11}$ renormalizes every component of the spin operator $\boldsymbol{S}_{i11}$, in agreement with Eq.~\eqn{RQ-B}.

\section{Entropy lower bound}
\label{sm:entropy}
\noindent
Within the variational free-energy approach of Eq.~\eqn{F:def}, the trial density matrix \eqn{density-matrix} yields a rigorous upper bound to the true free energy, $F\leq U(\rho_\text{gGut})-T\,S(\rho_\text{gGut})$. Since the entropy $S(\rho_\text{gGut})$ cannot be computed exactly within the GA, we replace it by a lower bound: as $T\geq0$, any lower bound on the entropy preserves the upper bound on $F$, and hence the variational character of the scheme. We now derive such a bound.\\
We consider the variational density matrix \eqn{density-matrix}, which we write here as
\begin{equation}
    \rho_\text{gGut} = \mathcal{P}\, \rho_\mathrm{qp}\, \mathcal{P}^\dagger
    = \prod_{i}\, \mathcal{P}_i\, \rho_\mathrm{qp}\, \mathcal{P}_i^\dagger\,,
    \label{rho-gGut-SM}
\end{equation}
where $\rho_\mathrm{qp}$ is a thermal non-interacting density matrix in the space of the auxiliary fermions $\mathbf{f}_i$, which are $M$-component spinors defined for each site $i$, while $\mathcal{P}_i$ is a linear map from the local auxiliary Hilbert space to the local one of the physical fermions $\mathbf{c}_i$, which are $L\leq M$-component spinors at each site $i$. We impose the two constraints
\begin{equation}
    \begin{aligned}
        \Tr\big(\rho_\mathrm{qp}\, \mathcal{P}_i^\dagger\, \mathcal{P}_i\big)
        &= \Tr\big(\rho_\mathrm{qp}\big) = 1\,, \\
        \Tr\big(\rho_\mathrm{qp}\, \mathcal{P}_i^\dagger\, \mathcal{P}_i\,
        \mathbf{f}_i\otimes\mathbf{f}_i^\dagger\big)
        &= \Tr\big(\rho_\mathrm{qp}\, \mathbf{f}_i\otimes\mathbf{f}_i^\dagger\big)\,.
    \end{aligned}
    \label{constraints-SM}
\end{equation}
Within the GA, assuming implicitly a lattice with infinite coordination number, it follows that
\begin{equation}
    \begin{aligned}
    \Tr\big(\rho_\text{gGut}\big) &= 1\,, \\
    \Tr\big(\rho_\text{gGut}\, A\big) &= \Tr\big(\rho_\mathrm{qp}\, A\big)\,,
    \end{aligned}
    \label{GA-trace-SM}
\end{equation}
with $A$ any one-body operator in the auxiliary $f$-space, and
\begin{equation}
    \begin{aligned}
    \Tr\big(\rho_\text{gGut}\, \mathcal{O}(\mathbf{c}_i^\dagger,\mathbf{c}_i)\big)
    &= \Tr\big(\rho_\mathrm{qp}\, \mathcal{P}_i^\dagger\, \mathcal{O}(\mathbf{c}_i^\dagger,\mathbf{c}_i)\, \mathcal{P}_i\big) \\
    &= \Tr\big(\rho_f(i)\, \mathcal{P}_i^\dagger\, \mathcal{O}(\mathbf{c}_i^\dagger,\mathbf{c}_i)\, \mathcal{P}_i\big)\,,
    \end{aligned}
\end{equation}
where $\rho_f(i) = \Tr_{j\neq i}\big(\rho_\mathrm{qp}\big)$. Defining $\mathcal{P}_i := \Phi_i\, \rho_f(i)^{-1/2}$ and $\mathcal{P}_i^\dagger := \rho_f(i)^{-1/2}\, \Phi_i^\dagger$, we obtain $\Tr\big(\rho_f(i)\, \mathcal{P}_i^\dagger\, \mathcal{O}\, \mathcal{P}_i\big) = \Tr\big(\Phi_i^\dagger\, \mathcal{O}\, \Phi_i\big)$.\\
The internal energy of the physical Hamiltonian is readily obtained within the GA; the main issue is to compute the entropy
\begin{equation}
    S(\rho_\text{gGut}) := -\Tr\big(\rho_\text{gGut}\, \ln\rho_\text{gGut}\big)\,.
\end{equation}
We define $\mathcal{Q}_i$, not to be confused with the spin renormalization factors $Q_{ab}$, such that
\begin{equation}
\mathcal{Q}_i = \mathcal{P}_i^\dagger\, \mathcal{P}_i \,,
\qquad
\mathcal{P}_i = U_i\sqrt{\mathcal{Q}_i} \, ,
\qquad 
\mathcal{P}_i^\dagger = \sqrt{\mathcal{Q}_i}\, U_i^\dagger
\end{equation} 
for unitary $U_i$, so that 
\begin{equation}
\mathcal{Q} = \prod_i \mathcal{Q}_i \,,
\qquad 
\mathcal{P} = U\sqrt{\mathcal{Q}}\, , 
\qquad
\mathcal{P}^\dagger = \sqrt{\mathcal{Q}}\, U^\dagger \,.
\end{equation}
It follows that
\beal
    S(\rho_\text{gGut}) &= -\Tr\Big(U\sqrt{\mathcal{Q}}\, \rho_\mathrm{qp}\, \sqrt{\mathcal{Q}}\, U^\dagger\, \ln\!\big(U\sqrt{\mathcal{Q}}\, \rho_\mathrm{qp}\, \sqrt{\mathcal{Q}}\, U^\dagger\big)\Big)\\
    &= -\Tr\Big(\sqrt{\mathcal{Q}}\, \rho_\mathrm{qp}\, \sqrt{\mathcal{Q}}\, \ln\!\big(\sqrt{\mathcal{Q}}\, \rho_\mathrm{qp}\, \sqrt{\mathcal{Q}}\big)\Big)\,.
\eal
A first lower bound follows from the Golden-Thompson-type trace inequality
\begin{equation}
-S(\rho_\text{gGut})=\Tr\Big(\rho_\text{gGut}\,\ln\big(\sqrt{\mathcal{Q}}\, \rho_\mathrm{qp}\, \sqrt{\mathcal{Q}}\big)\Big) \leq \Tr\Big(\rho_\text{gGut}\,\ln \rho_\mathrm{qp}\Big)
+ \Tr\Big(\rho_\text{gGut}\,\ln \mathcal{Q}\Big)
\end{equation}
which implies, through \eqn{GA-trace-SM} and since $\ln \rho_\mathrm{qp}$ is a one-body operator, that 
\beal
    S(\rho_\text{gGut}) &\geq -\Tr\big(\rho_\mathrm{qp}\, \mathcal{Q}\, \ln \rho_\mathrm{qp}\big) - \Tr\big(\mathcal{Q}^{1/2}\, \rho_\mathrm{qp}\, \mathcal{Q}^{1/2}\, \ln\mathcal{Q}\big)\\
    &= S(\rho_\mathrm{qp}) - \sum_i \Tr\Big(\sqrt{\mathcal{Q}_i}\, \rho_f(i)\, \sqrt{\mathcal{Q}_i}\, \ln\mathcal{Q}_i\Big)\\
    &= S(\rho_\mathrm{qp}) - \sum_i \Tr\big(\mathcal{Q}_i\, \rho_f(i)\, \ln\mathcal{Q}_i\big)\\
    &= S(\rho_\mathrm{qp}) - \sum_i \Tr\Big(\Phi_i^\dagger\Phi_i\, \ln\!\big(\Phi_i^\dagger\Phi_i\, \rho_f(i)^{-1}\big)\Big)\,.
    \label{bound 1}
\eal
However, this is a very poor bound since its right-hand side can become negative. To find a better loweer bound, we employ the following inequality~\cite{HiaiPetz1993} and still using \eqn{GA-trace-SM}, 
\begin{equation}
    S(\rho_\text{gGut}) - S(\rho_\mathrm{qp}) \geq - \frac{1}{p} \Tr\Big(\mathcal{Q}^{1/2}\, \rho_\mathrm{qp}\, \mathcal{Q}^{1/2}\, \ln\!\Big(\big(\mathcal{Q}^{1/2}\, \rho_\mathrm{qp}\, \mathcal{Q}^{1/2}\big)^p \big(\mathcal{Q}^{1/2}\, \rho_\mathrm{qp}\, \mathcal{Q}^{-1/2}\big)^{-p} \Big ) := - D_p(\rho_\mathrm{qp}, \mathcal{Q})\,.
    \label{bound 2}
\end{equation}
The inequality \eqn{bound 2} reduces to \eqn{bound 1} for $p=1$, while it becomes a strict equality as $p\to 0$. 
We emphasize that this lower bound becomes tighter as $p$ decreases from 1 to 0~\cite{HiaiPetz1993}. Furthermore, since $S(\rho_\text{gGut})< S(\rho_\mathrm{qp})$ and $D_p(\rho_\mathrm{qp}, \mathcal{Q})>0$, we can write
\begin{equation}
    D_p(\rho_\mathrm{qp}, \mathcal{Q}) := S(\rho_\mathrm{qp})\, W_p(\rho_\mathrm{qp},\mathcal{Q})\,,
\end{equation}
with $W_p(\rho_\mathrm{qp},\mathcal{Q}) > 0$, thus 
\begin{equation}
    S(\rho_\text{gGut}) \geq S(\rho_\mathrm{qp})\, \big(1 - W_p(\rho_\mathrm{qp},\mathcal{Q})\big)\,,\quad p\in[0,1]\,.
\end{equation}
For $T\to\infty$, $\rho_\mathrm{qp}\to\mathds{1}/d$, where $d$ is the dimension of the local auxiliary Hilbert space, so that 
\begin{equation}
\mathcal{Q}^{1/2}\rho_\mathrm{qp}\mathcal{Q}^{1/2}\to\mathcal{Q}/d \,,
\qquad
\mathcal{Q}^{1/2}\rho_\mathrm{qp}\mathcal{Q}^{-1/2}\to\mathds{1}/d \,.
\end{equation}
The argument of the logarithm in $D_p$ then simplifies to
\begin{equation}
    \big(\mathcal{Q}^{1/2}\rho_\mathrm{qp}\mathcal{Q}^{1/2}\big)^p \big(\mathcal{Q}^{1/2}\rho_\mathrm{qp}\mathcal{Q}^{-1/2}\big)^{-p}
    \;\to\; \left(\frac{\mathcal{Q}}{d}\right)^{\!p} \left(\frac{\mathds{1}}{d}\right)^{\!-p} = \mathcal{Q}^p\,,
\end{equation}
so that
\begin{equation}
    D_p(\rho_\mathrm{qp},\mathcal{Q}) \;\to\; \frac{1}{d}\,\Tr\!\left(\mathcal{Q}\ln\mathcal{Q}\right),
\end{equation}
which is independent of $p$, thus so is $W_p(\rho_\mathrm{qp},\mathcal{Q})$. Moreover, $W_p(\rho_\mathrm{qp},\mathcal{Q})$ decreases monotonically with decreasing $p$ from 1 to 0~\cite{HiaiPetz1993}, which implies
\begin{equation}
    0 < W_0(\rho_\mathrm{qp},\mathcal{Q}) \equiv \frac{S(\rho_\mathrm{qp}) - S(\rho_\text{gGut})}{S(\rho_\mathrm{qp})}
    \leq W_p(\rho_\mathrm{qp},\mathcal{Q}) \leq W_1(\rho_\mathrm{qp},\mathcal{Q})\,.
\end{equation}
On the other hand, since $S(\rho_\mathrm{qp})$ increases with $T$, being maximum and equal to $S_\infty$ at $T\to\infty$,
\begin{equation}
    W_1(\rho_\mathrm{qp},\mathcal{Q}) = \frac{D_1(\rho_\mathrm{qp},\mathcal{Q})}{S(\rho_\mathrm{qp})}
    \geq \frac{D_1(\rho_\mathrm{qp},\mathcal{Q})}{S_\infty}\,.
\end{equation}
We wish to demonstrate that there exists a $p$ with $0<p<1$ such that
\begin{equation}
    W_0(\rho_\mathrm{qp},\mathcal{Q}) \leq W_p(\rho_\mathrm{qp},\mathcal{Q}) = \frac{D_p(\rho_\mathrm{qp},\mathcal{Q})}{S(\rho_\mathrm{qp})}
    \leq \frac{D_1(\rho_\mathrm{qp},\mathcal{Q})}{S_\infty} \leq W_1(\rho_\mathrm{qp},\mathcal{Q})\,.
\end{equation}
We assume that $S(\rho_\mathrm{qp})\to 0$ for $T\to 0$, which implies $S(\rho_\text{gGut})\to 0$ as well. If both vanish in the same way, as expected when $\rho_\mathrm{qp}$ and $\rho_\text{gGut}$ describe the same low-energy physics, then $W_0(\rho_\mathrm{qp},\mathcal{Q})\to 0$, and the chain of inequalities
\begin{equation}
    W_0(\rho_\mathrm{qp},\mathcal{Q}) \equiv 0 \leq W_p(\rho_\mathrm{qp},\mathcal{Q})
    \leq \frac{D_1(\rho_\mathrm{qp},\mathcal{Q})}{S_\infty} \leq W_1(\rho_\mathrm{qp},\mathcal{Q})
    \label{chain-SM}
\end{equation}
holds strictly for $p$ sufficiently close to 0. Equation \eqn{chain-SM} remains valid even if
\begin{equation}
    \lim_{T\to 0}\, \frac{S(\rho_\mathrm{qp}) - S(\rho_\text{gGut})}{S(\rho_\mathrm{qp})} = C > 0\,,
\end{equation}
provided $C \leq D_1(\rho_\mathrm{qp},\mathcal{Q})/S_\infty$ for all $T$.\\
Let us summarize what the chain of inequalities \eqn{chain-SM} establishes. The exact entropy deficit of the variational state is encoded in $W_0(\rho_\mathrm{qp},\mathcal{Q}) = \big[S(\rho_\mathrm{qp})-S(\rho_\text{gGut})\big]/S(\rho_\mathrm{qp})$, which is not accessible within the GA. Equation~\eqn{chain-SM} shows that the single quantity
\begin{equation}
    \frac{D_1(\rho_\mathrm{qp},\mathcal{Q})}{S_\infty}
\end{equation}
is bounded from below and above by the exact deficit $W_0$ and the loose estimate $W_1$ respectively, i.e.\ $W_0 \leq D_1(\rho_\mathrm{qp},\mathcal{Q})/S_\infty \leq W_1$. Because it lies above $W_0$, replacing the true deficit by $D_1/S_\infty$ can only underestimate the entropy, so it provides a lower bound. Crucially, $D_1/S_\infty$ depends only on the saddle-point objects $\mathcal{Q}$ and $\rho_\mathrm{qp}$ and on the dimension $d$ of the local auxiliary Hilbert space, so it is fully computable within the GA at no extra cost.\\
\noindent
Identifying the infinite-temperature entropy with $S_\infty = N\ln d$, with $N$ the number of sites, and using $D_1(\rho_\mathrm{qp},\mathcal{Q}) = \Tr\big(\rho_\mathrm{qp}\, \mathcal{Q}\, \ln\mathcal{Q}\big)$, we therefore arrive at
\begin{equation}
    S(\rho_\text{gGut}) \geq S(\rho_\mathrm{qp})\,\Big(1 - \frac{\Tr\big(\rho_\mathrm{qp}\, \mathcal{Q}\, \ln\mathcal{Q}\big)}{N\ln d}\Big)\,,
\end{equation}
which is precisely Eq.~\eqn{lower bound}. Inserted into $F\leq U(\rho_\text{gGut})-T\,S(\rho_\text{gGut})$, it preserves the upper bound on the free energy and hence the variational character of the finite-temperature ghost-Gutzwiller scheme.

\subsection{Explicit evaluation in the two limiting cases}
\label{sm:entropy-cases}
\noindent
We now evaluate the two ingredients entering the entropy bound, the distortion term 
\begin{equation}
    D_1(\rho_\mathrm{qp},\mathcal{Q})=\Tr\big(\rho_\mathrm{qp}\,\mathcal{Q}\,\ln\mathcal{Q}\big)=\sum_i\Tr\big(\Phi_i^\dagger\Phi_i\,\ln(\Phi_i^\dagger\Phi_i\,\rho_f(i)^{-1})\big)
\end{equation}
and the infinite temperature limit of the quasiparticles entropy $S_\infty=\ln d$, for the two cases A/B of the main text. Both quantities are extensive, and we quote them per site; the estimate involves only their intensive ratio. In both cases we work in the local Fock basis $\{0,\up,\down,2\}$ of the auxiliary orbitals, and we use the impurity amplitudes of Eqs.~\eqn{psi-A}--\eqn{psi-B} together with the auxiliary occupations quoted in Sec.~II.

\subsubsection{Conventional Gutzwiller solution (case A)}
\noindent
In this case orbital $2$ is doubly occupied and factorizes out, leaving the four-dimensional local space of orbital $1$, whose reduced density matrix and variational matrix are both diagonal,
\begin{equation}
    \begin{aligned}
    \rho_f &= \tfrac{1}{4}\,\mathrm{diag}\big((1+\delta)^2,\,1-\delta^2,\,1-\delta^2,\,(1-\delta)^2\big)\,, \\
    \Phi^\dagger\Phi &= \mathrm{diag}\Big(\delta,\,\tfrac{1-\delta}{2},\,\tfrac{1-\delta}{2},\,0\Big)\,.
    \end{aligned}
\end{equation}
The distortion term and the infinite-temperature entropy per site thus read
\begin{equation}
    \begin{aligned}
    \frac{D_1}{N} 
    &= \delta\,\ln\frac{4\delta}{(1+\delta)^2} - (1-\delta)\,\ln\frac{1+\delta}{2}\,, \\
    \frac{S_\infty}{N} &= -(1-\delta)\,\ln\frac{1-\delta}{2} - (1+\delta)\,\ln\frac{1+\delta}{2}\,,
    \end{aligned}
\end{equation}
where $S_\infty/N$ is the entropy per site of $2N$ auxiliary spin-orbitals hosting $(1-\delta)N$ fermions, obtained from the binomial count in the Stirling limit.

\subsubsection{Fractionalized $\mathrm{FL}^{\!*}$ solution (case B)}
\noindent
Now both orbitals are active: orbital $1$ is half-filled and orbital $2$ hosts $2-\delta$ electrons. In the four-dimensional subspace $\{(\up,2),(\down,2),(\up,\down),(\down,\up)\}$ the variational matrix acquires a non-diagonal singlet block,
\begin{equation}
    \Phi^\dagger\Phi
    = \frac{1-\delta}{2}\,\big(\ket{\up,2}\bra{\up,2}+\ket{\down,2}\bra{\down,2}\big)
    + \frac{\delta}{2}\begin{pmatrix}1&-1\\-1&1\end{pmatrix}_{\!\{(\up,\down),(\down,\up)\}}\,,
\end{equation}
while the reduced density matrix stays diagonal, with entries 
\begin{equation}
    \begin{aligned}
        \rho_f(\up,2)&=\rho_f(\down,2)=(2-\delta)^2/16 \,, \\
        \rho_f(\up,\down)&=\rho_f(\down,\up)=(2-\delta)\delta/16 \,.
    \end{aligned}
\end{equation}
Since $\rho_f$ is proportional to the identity within the singlet block, the block contributes only through the non-zero eigenvalue $\delta$ of $\Phi^\dagger\Phi$, giving
\begin{equation}
    \begin{aligned}
    \frac{D_1}{N} &= (1-\delta)\,\ln\frac{2(1-\delta)}{(1-\delta/2)^2} + \delta\,\ln\frac{8}{1-\delta/2}\,, \\
    \frac{S_\infty}{N} &= \ln 4 - (2-\delta)\,\ln\frac{2-\delta}{2} - \delta\,\ln\frac{\delta}{2}\,,
    \end{aligned}
\end{equation}
the latter being the per-site entropy of a half-filled orbital plus an orbital at filling $2-\delta$.

\subsubsection{High-temperature consistency}
\noindent
We can explicitly verify that for both solutions the estimate saturates correctly as $T\to\infty$: subtracting the distortion term from $S_\infty$ recovers the exact infinite-temperature entropy of the $t$-$J$ model at doping $\delta$,
\begin{equation}
    \frac{S_\infty-D_1}{N} = -\delta\,\ln\delta - (1-\delta)\,\ln\frac{1-\delta}{2}\,,
\end{equation}
i.e.\ the configurational entropy of $\delta N$ empty and $(1-\delta)N$ singly-occupied sites. The identity holds for case A and case B alike, confirming that the $D_1/S_\infty$ estimate is exact at high temperature in both the conventional and the fractionalized phase.

\section{Finite magnetization and uniform spin susceptibility}
\label{sm:magnetization}
\noindent
In this section we extend the construction of the main text to states with a finite uniform magnetization along the $z$-axis, which is needed to compute the static uniform spin susceptibility $\chi$. We first introduce the magnetized counterparts of the impurity wavefunctions \eqn{psi-A} and \eqn{psi-B}. Since $SU(2)$ symmetry is now broken, the transverse and longitudinal components of the spin operator have to be projected separately, and the projection of $S^z$ generates additional density vertices. We then write the variational free energy at fixed magnetization, describe its self-consistent optimization, and finally show how $\chi$ is extracted from its curvature.\\
\noindent
Throughout this section we fix the magnetization per site
\begin{equation}
    m = \langle n_{i\up} - n_{i\down}\rangle = 2\,m_z\,,\qquad |m|\leq 1-\delta\,,
\end{equation}
with $n_{i\sigma}=c^\dagger_{i\sigma}c^\dagga_{i\sigma}$ and $m_z=\langle S^z_i\rangle$ as in the main text, and, whenever $\sigma$ appears as a number, $\sigma=\pm1$ for $\up,\down$. All derivatives will be taken at fixed doping $\delta$. We define the spin-resolved physical single-occupancy probabilities
\begin{equation}
    p_\sigma = \frac{1-\delta+\sigma m}{2}\,,
\end{equation}
which reduce to $\cos^2\theta/2$ of Eqs.~\eqn{psi-A} and \eqn{psi-B} at $m=0$.

\subsection{Magnetized impurity wavefunctions}
\noindent
\textit{Case A.} The natural extension of Eq.~\eqn{psi-A} is
\begin{equation}
    \ket{\psi^\mathrm{A}_\mathrm{imp}(m)} = \sqrt{p_\up}\,\ket{\up;\down,0} - \sqrt{p_\down}\,\ket{\down;\up,0} + \sqrt{\delta}\,\ket{0;2,0}\,,
    \label{psi-A-m}
\end{equation}
which has physical density $1-\delta$ and magnetization $m$. The GA constraint gives $n^\mathrm{qp}_{1\sigma}=p_\sigma$ and $n^\mathrm{qp}_{2\sigma}=1$: the auxiliary orbital 1 carries exactly the physical density and magnetization, while orbital 2 remains a fully occupied spectator. At fixed $(\delta,m)$ there is no free local parameter.\\
\noindent
\textit{Case B.} We extend Eq.~\eqn{psi-B} as
\begin{equation}
    \ket{\psi^\mathrm{B}_\mathrm{imp}(m,\eta)} = \sqrt{p_\up}\,\ket{\up;\down,0} - \sqrt{p_\down}\,\ket{\down;\up,0} + \sqrt{\frac{\delta+\eta}{2}}\,\ket{0;\up,\down} - \sqrt{\frac{\delta-\eta}{2}}\,\ket{0;\down,\up}\,,
    \qquad |\eta|\leq\delta\,.
    \label{psi-B-m}
\end{equation}
The parameter $\eta$ only reweights the two bath configurations associated with the empty impurity, hence it affects neither the physical density nor the physical magnetization. It does, however, redistribute the auxiliary magnetization between the two orbitals. Indeed, the GA constraint now gives
\begin{equation}
    n^\mathrm{qp}_{1\sigma} = x_\sigma \equiv \frac{1+\sigma m_1}{2}\,,\qquad
    n^\mathrm{qp}_{2\sigma} = y_\sigma \equiv \frac{2-\delta+\sigma\eta}{2}\,,\qquad
    m_1 \equiv m-\eta\,,
    \label{qp-densities-B-m}
\end{equation}
so that the spinon orbital 1 carries magnetization $m_1$ and the pocket orbital 2 magnetization $\eta$, with $m_1+\eta=m$. We further define $w_\sigma\equiv 1-y_\sigma=(\delta-\sigma\eta)/2$. In the following, $\eta$ is an additional variational parameter, optimized at each $m$ and $T$.

\subsection{Projected operators}
\noindent
Proceeding as in Sec.~II, the local operators read
\begin{equation}
    \begin{aligned}
    \mathcal{P}_\mathrm{A} &= \sum_\sigma \lambda_\sigma\,\ket{\sigma}\bra{\sigma,2} + \lambda_0\,\ket{0}\bra{0,2}\,,\\
    \mathcal{P}_\mathrm{B} &= \sum_\sigma \lambda_\sigma\,\ket{\sigma}\bra{\sigma,2} + \lambda'_\up\,\ket{0}\bra{\up,\down} - \lambda'_\down\,\ket{0}\bra{\down,\up}\,,
    \end{aligned}
\end{equation}
where $\lambda_\sigma=\sqrt{p_\sigma/P_0(\sigma,2)}$, $\lambda_0=\sqrt{\delta/P_0(0,2)}$, and $\lambda'_\sigma=\sqrt{w_\sigma/P_0(\sigma,\bar\sigma)}$, with
\begin{equation}
    \begin{aligned}
    &\text{case A:}\quad P_0(\sigma,2)=p_\sigma\,(1-p_{\bar\sigma})\,,\qquad P_0(0,2)=(1-p_\up)(1-p_\down)\,,\\
    &\text{case B:}\quad P_0(\sigma,2)=x_\sigma^2\,y_\up y_\down\,,\qquad\;\;\; P_0(\sigma,\bar\sigma)=x_\sigma^2\,y_{\bar\sigma}\,w_\sigma\,.
    \end{aligned}
\end{equation}
In case B we used $1-x_{\bar\sigma}=x_\sigma$. Note that, at finite $m$, $\lambda'_\up\neq\lambda'_\down$.\\
\noindent
\textit{Hopping.} The projection of $c^\dagger_\sigma$ follows exactly Sec.~II, the only difference being that the spectator densities are now replaced by their spin-resolved averages. The spin-flip term of case B still gives no contribution, since $\langle f^\dagger_{1\sigma}f^\dagga_{1\bar\sigma}\rangle_\mathrm{qp}=0$ for a magnetization along $z$. We obtain $\mathcal{P}^\dagger c^\dagger_\sigma\mathcal{P}\to R_{a\sigma}\,f^\dagger_{a\sigma}$, with the hopping carried by orbital $a=1$ in case A and $a=2$ in case B, and spin-dependent renormalization factors
\begin{equation}
    \text{case A:}\quad R_{1\sigma}^2=\frac{\delta}{1-p_\sigma}=\frac{2\delta}{1+\delta-\sigma m}\,,\qquad\quad
    \text{case B:}\quad R_{2\sigma}^2=\frac{p_\sigma}{y_\sigma}=\frac{1-\delta+\sigma m}{2-\delta+\sigma\eta}\,.
    \label{R-m}
\end{equation}
\textit{Transverse spin.} As in Sec.~II, $\mathcal{P}^\dagger S^+\mathcal{P}=\lambda_\up\lambda_\down\,S^+_{i11}\,n_{2\up}n_{2\down}$, so that $\mathcal{P}^\dagger S^\pm\mathcal{P}\to Q_\perp\,S^\pm_{i11}$, with
\begin{equation}
    \text{case A:}\quad Q_\perp=\frac{2}{\sqrt{(1+\delta)^2-m^2}}\,,\qquad\quad
    \text{case B:}\quad Q_\perp=\frac{\sqrt{p_\up p_\down}}{x_\up x_\down}=\frac{2\sqrt{(1-\delta)^2-m^2}}{1-m_1^2}\,.
\end{equation}
\textit{Longitudinal spin.} For $S^z$, the projection gives the exact local identity
\begin{equation}
    \mathcal{P}^\dagger S^z\mathcal{P} = \frac12\sum_\sigma \sigma\,\lambda_\sigma^2\,\ket{\sigma,2}\bra{\sigma,2}
    = \frac12\sum_\sigma \sigma\,\lambda_\sigma^2\; n_{1\sigma}\,(1-n_{1\bar\sigma})\,n_{2\up}\,n_{2\down}\,.
\end{equation}
Within the GA, we expand it around the averages, $\delta n_{a\sigma}=n_{a\sigma}-n^\mathrm{qp}_{a\sigma}$, retaining the constant and the terms linear in $\delta n_{a\sigma}$, i.e., the vertices with at most two fermion legs, consistently with the treatment of $c^\dagger_\sigma$ and $S^\pm$. This yields
\begin{equation}
    \mathcal{P}^\dagger S^z_i\,\mathcal{P} \to \frac{m}{2} + \sum_\sigma \sigma\, q_\sigma\, \delta n_{i1\sigma} + \sum_\sigma v_\sigma\, \delta n_{i2\sigma}\,,
    \label{Sz-mapping}
\end{equation}
with
\begin{equation}
    \begin{aligned}
    &\text{case A:}\quad q_\sigma=\frac{1-\sigma m}{1+\delta-\sigma m}\,,\qquad v_\sigma=0\,,\\
    &\text{case B:}\quad q_\sigma=\frac12\sum_{\sigma'}\frac{p_{\sigma'}}{x_{\sigma'}}=\frac{1-\delta-m\,m_1}{1-m_1^2}\,,\qquad v_\sigma=\frac{m}{2\,y_\sigma}\,.
    \end{aligned}
    \label{q-v}
\end{equation}
A few remarks are in order. First, the constant term in \eqn{Sz-mapping} is not renormalized: $\langle\mathcal{P}^\dagger S^z\mathcal{P}\rangle = m/2$ exactly, for any $\eta$. Second, at $m=0$ we have $q_\sigma=Q_\perp/2=Q_{11}/2$ and $v_\sigma=0$, so that \eqn{Sz-mapping} reduces to the isotropic mapping \eqn{mapping} with the factors \eqn{RQ-A} and \eqn{RQ-B}. Third, at finite $m$ the longitudinal mapping is no longer exhausted by a spin vertex: in case A, $q_\up\neq q_\down$, i.e., $S^z$ also couples to the charge fluctuations of orbital 1; in case B, $S^z$ couples to the density fluctuations of the pocket orbital 2 through $v_\sigma\propto m$.

\subsection{Variational energy}
\noindent
We take $\ket{\psi_\mathrm{qp}}$ uniform, inversion symmetric, and conserving $S^z$: orbital 1 is described by a spin-polarized BCS state with real singlet $d$-wave pairing, while in case B orbital 2 is a normal metal and $\ket{\psi_\mathrm{qp}}=\ket{\psi_1}\ket{\psi_2}$, as in the main text. With $f^\dagga_{ia\sigma}=N^{-1/2}\sum_\bk \esp{i\bk\cdot\bR_i}\,f^\dagga_{a\bk\sigma}$, $N$ the number of sites, and $\hat\alpha$ the unit lattice vector along $\alpha=x,y$, we denote by
\begin{equation}
    \begin{aligned}
    n_{a\bk\sigma} &= \langle f^\dagger_{a\bk\sigma}f^\dagga_{a\bk\sigma}\rangle\,, &
    \Delta_\bk &= \langle f^\dagger_{1\bk\up}f^\dagger_{1-\bk\down}\rangle\,,\\
    K_{a\alpha\sigma} &= \langle f^\dagger_{ia\sigma}f^\dagga_{i+\hat\alpha,a\sigma}\rangle = \frac1N\sum_\bk \cos k_\alpha\, n_{a\bk\sigma}\,, &\quad
    \Delta_\alpha &= \langle f^\dagger_{i1\up}f^\dagger_{i+\hat\alpha,1\down}\rangle = \frac1N\sum_\bk \cos k_\alpha\,\Delta_\bk\,,
    \end{aligned}
\end{equation}
the momentum-space averages and the corresponding nearest-neighbor bond amplitudes, with $K_{ax\sigma}=K_{ay\sigma}$ and $\Delta_x=-\Delta_y$ for $d$-wave symmetry. Substituting \eqn{R-m}--\eqn{q-v} into the Hamiltonian \eqn{Ham} and applying Wick's theorem, the energy per site reads
\begin{equation}
    \begin{aligned}
    E(m,\eta) &= \frac1N\sum_{\bk\sigma}R^2_{a\sigma}\,\epsilon_\bk\, n_{a\bk\sigma} + \frac{J m^2}{2}\\
    &\quad - J\sum_{\alpha}\Big[\sum_\sigma q_\sigma^2\,K_{1\alpha\sigma}^2 + Q_\perp^2\,K_{1\alpha\up}K_{1\alpha\down} + g\,\Delta_\alpha^2 + \sum_\sigma v_\sigma^2\,K_{2\alpha\sigma}^2\Big]\,,
    \qquad g = 2\,q_\up q_\down + Q_\perp^2\,,
    \end{aligned}
    \label{E-m}
\end{equation}
where $\epsilon_\bk=-2t(\cos k_x+\cos k_y)-4t'\cos k_x\cos k_y$ is the bare dispersion, with $t$ and $t'$ the nearest- and next-nearest-neighbor hopping amplitudes of \eqn{Ham}, $\sum_\alpha$ runs over $\alpha=x,y$, and $a=1$ ($a=2$) in case A (B). Case A has no $\eta$ dependence and no pocket term. The second term is the physical Hartree energy, $J(m/2)^2$ per bond with two bonds per site, and it comes solely from the unrenormalized constant in \eqn{Sz-mapping}: the terms linear in $\delta n_{a\sigma}$ have vanishing average because of the GA constraints, so that the Hartree contractions of the auxiliary operators drop out and only their connected (Fock and anomalous) contractions survive. In case B, the mixed spinon-pocket terms vanish as $\ket{\psi_\mathrm{qp}}$ is a product state. At $m=0$, \eqn{E-m} reduces to the Hartree-Fock decoupling of Eqs.~\eqn{Ham-A} and \eqn{Ham-B}.

\subsection{Self-consistency and free energy}
\noindent
Since \eqn{E-m} is a quadratic function of the averages $n_{a\bk\sigma}$ and $\Delta_\bk$, its constrained minimization over Fock states is a Hartree-Fock-Bogoliubov problem. Defining $\xi_{a\bk\sigma}\equiv\partial (N E)/\partial n_{a\bk\sigma}-\mu_a-\sigma h_a$ and $-2\mathcal{D}_\bk\equiv\partial (N E)/\partial\Delta_\bk$, where $\mu_a$ and $h_a$ are Lagrange multipliers, the optimal $\rho_\mathrm{qp}$ is the ground state, or the thermal state at finite temperature (see below), of
\begin{equation}
    H_\mathrm{MF} = \sum_{\bk\sigma}\xi_{1\bk\sigma}\,f^\dagger_{1\bk\sigma}f^\dagga_{1\bk\sigma} - \sum_\bk\Big(\mathcal{D}_\bk\,f^\dagger_{1\bk\up}f^\dagger_{1-\bk\down} + \text{H.c.}\Big) + \sum_{\bk\sigma}\xi_{2\bk\sigma}\,f^\dagger_{2\bk\sigma}f^\dagga_{2\bk\sigma}\,,
\end{equation}
where the last term is present only in case B, and
\begin{equation}
    \begin{aligned}
    \xi_{1\bk\sigma} &= \big[R^2_{1\sigma}\,\epsilon_\bk\big]_\mathrm{A} - J\sum_\alpha\big(2q_\sigma^2\,K_{1\alpha\sigma} + Q_\perp^2\,K_{1\alpha\bar\sigma}\big)\cos k_\alpha - \mu_1 - \sigma h_1\,,\\
    \mathcal{D}_\bk &= J\,g\,\Delta_x\,(\cos k_x - \cos k_y)\,,\\
    \xi_{2\bk\sigma} &= R^2_{2\sigma}\,\epsilon_\bk - 2J\,v_\sigma^2\sum_\alpha K_{2\alpha\sigma}\cos k_\alpha - \mu_2 - \sigma h_2\,,
    \end{aligned}
\end{equation}
the bare dispersion in $\xi_{1\bk\sigma}$ being present only in case A. The multipliers $\mu_a$ and $h_a$ enforce the GA constraints, $N^{-1}\sum_\bk n_{1\bk\sigma}=p_\sigma$ in case A, and $N^{-1}\sum_\bk n_{1\bk\sigma}=x_\sigma$, $N^{-1}\sum_\bk n_{2\bk\sigma}=y_\sigma$ in case B. We stress that $h_a$ are not the physical magnetic field. We note that the spin-averaged spinon hopping and the pairing of case B share the same coupling $g$ at any $m$, so that the $SU(2)$ gauge redundancy discussed in the main text survives and we can keep working with the $d$-wave representative.\\
\noindent
The spinon part is diagonalized by a Bogoliubov transformation. With
\begin{equation}
    \bar\xi_\bk=\frac{\xi_{1\bk\up}+\xi_{1\bk\down}}{2}\,,\qquad
    \zeta_\bk=\frac{\xi_{1\bk\up}-\xi_{1\bk\down}}{2}\,,\qquad
    E_\bk=\sqrt{\bar\xi_\bk^2+\mathcal{D}_\bk^2}\,,\qquad
    u_\bk^2, v_\bk^2 = \frac12\Big(1\pm\frac{\bar\xi_\bk}{E_\bk}\Big)\,,
\end{equation}
the quasiparticle energies are $E_{\bk\sigma}=E_\bk+\sigma\zeta_\bk$: the spin splitting $\zeta_\bk$ only shifts the two branches, leaving the coherence factors unchanged. At temperature $\tau$, specified below, the Bogoliubov quasiparticles have occupations $\nu_{\bk\sigma}=n_F(E_{\bk\sigma})$, with $n_F(E)=1/(\esp{E/\tau}+1)$, and
\begin{equation}
    n_{1\bk\sigma} = u_\bk^2\,\nu_{\bk\sigma} + v_\bk^2\,\big(1-\nu_{\bk\bar\sigma}\big)\,,\qquad
    \Delta_\bk = \frac{\mathcal{D}_\bk}{2E_\bk}\,\big(1-\nu_{\bk\up}-\nu_{\bk\down}\big)\,,
    \label{BCS-m}
\end{equation}
and $n_{2\bk\sigma}=n_F(\xi_{2\bk\sigma})$. It follows that $N^{-1}\sum_\bk(n_{1\bk\up}-n_{1\bk\down})=N^{-1}\sum_\bk(\nu_{\bk\up}-\nu_{\bk\down})$: the spinons can only be magnetized by thermally or energetically occupied Bogoliubov quasiparticles, which in turn Pauli-block the pairing. The bond amplitudes $K_{a\alpha\sigma}$ and $\Delta_\alpha$ computed from \eqn{BCS-m} are fed back into $H_\mathrm{MF}$ until convergence. The energy is then evaluated from the functional \eqn{E-m}, and not from the eigenvalues of $H_\mathrm{MF}$, which would require the subtraction of double-counting constants.\\
\noindent
At finite temperature, the free energy per site at fixed $(m,\eta)$ reads, using the entropy bound \eqn{lower bound},
\begin{equation}
    F(m,\eta) = \min_{\rho_\mathrm{qp}}\Big[E(m,\eta) - T\,\varphi(m,\eta)\, s_\mathrm{qp}\Big]\,,\qquad
    \varphi(m,\eta)=1-\frac{D_1}{S_\infty}\,,
    \label{F-m}
\end{equation}
where $s_\mathrm{qp}=N^{-1}\sum_{\bk\sigma}\big[s(\nu_{\bk\sigma}) + s(n_{2\bk\sigma})\big]$ is the entropy per site of $\rho_\mathrm{qp}$, with $s(z)=-z\ln z-(1-z)\ln(1-z)$ and the pocket term present only in case B. Since $D_1$ and $S_\infty$ depend only on the local density matrices, which are fixed by the constraints, $\varphi$ does not depend on $\rho_\mathrm{qp}$. The minimization in \eqn{F-m} therefore yields the self-consistent solution above at the effective temperature $\tau=\varphi\,T$. Repeating the steps of Sec.~III~A, we find for case A, where, in the Fock basis $\{\ket{0},\ket{\up},\ket{\down},\ket{2}\}$ of orbital 1, both $\Phi^\dagger\Phi=\mathrm{diag}(\delta,p_\up,p_\down,0)$ and $\rho_f=\mathrm{diag}\big(P_0(0,2),P_0(\up,2),P_0(\down,2),p_\up p_\down\big)$ are diagonal,
\begin{equation}
    \frac{D_1}{N}=\delta\ln\delta-\sum_\sigma(1-p_\sigma)\ln(1-p_\sigma)\,,\qquad
    \frac{S_\infty}{N}=\sum_\sigma s(p_\sigma)\,.
\end{equation}
In case B, $\Phi^\dagger\Phi$ has weights $p_\sigma$ on $\ket{\sigma,2}$ and, in the block $\{\ket{\up,\down},\ket{\down,\up}\}$, it is the rank-one matrix $\boldsymbol{u}\,\boldsymbol{u}^T$ with $\boldsymbol{u}=(\sqrt{w_\up},-\sqrt{w_\down})$, while $\rho_f$ is diagonal with entries $P_0(\sigma,2)$ and $P_0(\sigma,\bar\sigma)$. Hence
\begin{equation}
    \frac{D_1}{N}=\sum_\sigma p_\sigma\ln\frac{p_\sigma}{x_\sigma^2\,y_\up y_\down} + \delta\,\ln\Big[\sum_\sigma\frac{1}{x_\sigma^2\,y_{\bar\sigma}}\Big]\,,\qquad
    \frac{S_\infty}{N}=\sum_\sigma\big[s(x_\sigma)+s(y_\sigma)\big]\,.
\end{equation}
Both expressions reduce to those of Sec.~III~A at $m=\eta=0$. As $T\to\infty$, the exact entropy of the $t$-$J$ model at doping $\delta$ and magnetization $m$ is $S_\mathrm{exact}/N=-\delta\ln\delta-\sum_\sigma p_\sigma\ln p_\sigma$. In case A, $(S_\infty-D_1)/N=S_\mathrm{exact}/N$ for any $m$. In case B, instead,
\begin{equation}
    \frac{S_\infty-D_1}{N}=\frac{S_\mathrm{exact}}{N}-\delta\,\bigg[\ln\Big(\sum_\sigma\frac{w_\sigma}{\delta}\,\frac{1}{P_0(\sigma,\bar\sigma)}\Big)-\sum_\sigma\frac{w_\sigma}{\delta}\,\ln\frac{1}{P_0(\sigma,\bar\sigma)}\bigg]\,.
\end{equation}
Since $w_\up+w_\down=\delta$, the weights $w_\sigma/\delta$ sum to one, and by concavity of the logarithm the square bracket is non-negative. It vanishes only when $P_0(\up,\down)=P_0(\down,\up)$. Unlike in case A, this condition is not automatically satisfied at finite $m$. However, for any $m$ there exists an $\eta$ that fulfills it, so that the optimization over $\eta$, which as $T\to\infty$ amounts to maximizing $S_\infty-D_1$, recovers the exact high-temperature entropy and hence the exact Curie law $\chi\to(1-\delta)/(4T)$.

\subsection{Uniform spin susceptibility}
\noindent
We add to the Hamiltonian \eqn{Ham} the Zeeman term $-h\sum_i S^z_i$. Since $\langle\mathcal{P}^\dagger S^z\mathcal{P}\rangle=m/2$ exactly, see \eqn{Sz-mapping}, the Zeeman energy per site is $-h\,m/2$ and depends on the variational state only through $m$. Defining the optimized free energy at fixed magnetization,
\begin{equation}
    F^\star(m)=\min_\eta\,F(m,\eta)\,,
\end{equation}
with no minimization over $\eta$ in case A, the free energy in the presence of the field is $\min_m\big[F^\star(m)-h\,m/2\big]$. Expanding $F^\star(m)=F^\star(0)+\kappa\,m^2/2+\dots$, stationarity gives $\kappa\,m=h/2$ to linear order, hence
\begin{equation}
    \chi=\frac{\partial m_z}{\partial h}\Big|_{h=0}=\frac{1}{4\kappa}=\bigg(\frac{\partial^2F^\star}{\partial m_z^2}\Big|_{m_z=0}\bigg)^{-1}\,,\qquad
    \kappa=\frac{\partial^2F^\star}{\partial m^2}\Big|_{m=0}\,.
    \label{chi-def}
\end{equation}
In practice, $F^\star(m)$ is computed exactly as in the paramagnetic case, the only difference being that, instead of fixing the total density $n=1-\delta$, we fix separately the two spin densities $n_\sigma=p_\sigma=(1-\delta+\sigma m)/2$. The curvature $\kappa$ is then obtained by finite differences of $F^\star(m)$ around $m=0$.

\section{Dependence of the phase diagram on the model parameters}
\label{sm:parameters}
\begin{figure}[h]
    \centering
    \includegraphics[width=\textwidth]{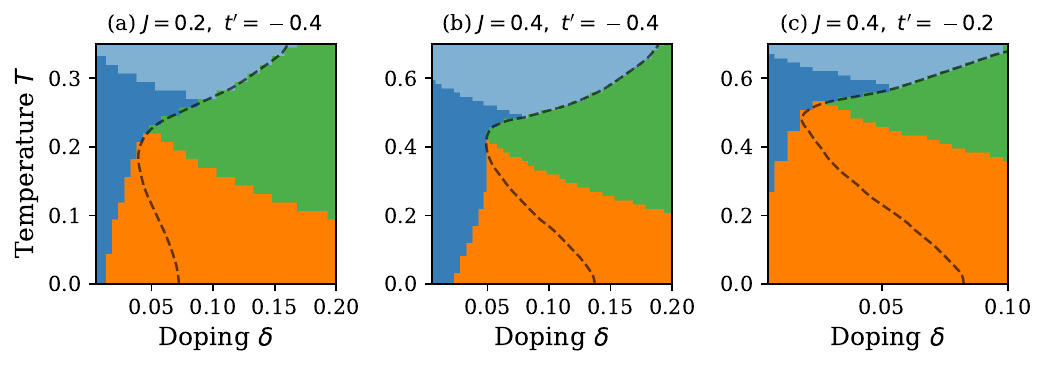}
    \caption{Temperature--doping phase diagram of the $t$-$J$ model within the ghost-Gutzwiller Ansatz for (a) $J/t=0.2$, $t^\prime/t=-0.4$; (b) $J/t=0.4$, $t^\prime/t=-0.4$, as in Fig.~\ref{fig:PhaseDiagram}; (c) $J/t=0.4$, $t^\prime/t=-0.2$. Colors and lines as in Fig.~\ref{fig:PhaseDiagram}. Note the different temperature scale in panel (a).}
    \label{fig:PD-parameters}
\end{figure}

\noindent
In this section we assess how the phase diagram of Fig.~\ref{fig:PhaseDiagram} depends on the choice of the parameters in \eqn{Ham}. In Fig.~\ref{fig:PD-parameters} we compare the phase diagram for $J/t=0.4$, $t^\prime/t=-0.4$, used in the main text, with those obtained by reducing either the exchange coupling, $J/t=0.2$, or the magnitude of the next-nearest-neighbor hopping, $t^\prime/t=-0.2$. In all cases we find the same sequence of phases: a $\mathrm{FL}^{\!*}$ at low doping, with a $\pi$-flux order parameter finite below $T^{\!*}$, a conventional FL with a large Fermi surface at high doping, and a $d$-wave superconducting dome intruding between them. Changing the parameters only shifts the location of the transition lines. In particular, the overall temperature scale, including $T^{\!*}$ and the superconducting critical temperature, is set by $J$, while reducing $|t^\prime|$ shrinks the doping range over which the $\mathrm{FL}^{\!*}$ is stable.

\begin{thebibliography}{70}%
\makeatletter
\providecommand \@ifxundefined [1]{%
 \@ifx{#1\undefined}
}%
\providecommand \@ifnum [1]{%
 \ifnum #1\expandafter \@firstoftwo
 \else \expandafter \@secondoftwo
 \fi
}%
\providecommand \@ifx [1]{%
 \ifx #1\expandafter \@firstoftwo
 \else \expandafter \@secondoftwo
 \fi
}%
\providecommand \natexlab [1]{#1}%
\providecommand \enquote  [1]{``#1''}%
\providecommand \bibnamefont  [1]{#1}%
\providecommand \bibfnamefont [1]{#1}%
\providecommand \citenamefont [1]{#1}%
\providecommand \href@noop [0]{\@secondoftwo}%
\providecommand \href [0]{\begingroup \@sanitize@url \@href}%
\providecommand \@href[1]{\@@startlink{#1}\@@href}%
\providecommand \@@href[1]{\endgroup#1\@@endlink}%
\providecommand \@sanitize@url [0]{\catcode `\\12\catcode `\$12\catcode
  `\&12\catcode `\#12\catcode `\^12\catcode `\_12\catcode `\%12\relax}%
\providecommand \@@startlink[1]{}%
\providecommand \@@endlink[0]{}%
\providecommand \url  [0]{\begingroup\@sanitize@url \@url }%
\providecommand \@url [1]{\endgroup\@href {#1}{\urlprefix }}%
\providecommand \urlprefix  [0]{URL }%
\providecommand \Eprint [0]{\href }%
\providecommand \doibase [0]{https://doi.org/}%
\providecommand \selectlanguage [0]{\@gobble}%
\providecommand \bibinfo  [0]{\@secondoftwo}%
\providecommand \bibfield  [0]{\@secondoftwo}%
\providecommand \translation [1]{[#1]}%
\providecommand \BibitemOpen [0]{}%
\providecommand \bibitemStop [0]{}%
\providecommand \bibitemNoStop [0]{.\EOS\space}%
\providecommand \EOS [0]{\spacefactor3000\relax}%
\providecommand \BibitemShut  [1]{\csname bibitem#1\endcsname}%
\let\auto@bib@innerbib\@empty
\bibitem [{\citenamefont {Norman}\ \emph {et~al.}(1998)\citenamefont {Norman},
  \citenamefont {Ding}, \citenamefont {Randeria}, \citenamefont {Campuzano},
  \citenamefont {Yokoya}, \citenamefont {Takeuchi}, \citenamefont {Takahashi},
  \citenamefont {Mochiku}, \citenamefont {Kadowaki}, \citenamefont
  {Guptasarma},\ and\ \citenamefont {Hinks}}]{Norman1998}%
  \BibitemOpen
  \bibfield  {author} {\bibinfo {author} {\bibfnamefont {M.~R.}\ \bibnamefont
  {Norman}}, \bibinfo {author} {\bibfnamefont {H.}~\bibnamefont {Ding}},
  \bibinfo {author} {\bibfnamefont {M.}~\bibnamefont {Randeria}}, \bibinfo
  {author} {\bibfnamefont {J.~C.}\ \bibnamefont {Campuzano}}, \bibinfo {author}
  {\bibfnamefont {T.}~\bibnamefont {Yokoya}}, \bibinfo {author} {\bibfnamefont
  {T.}~\bibnamefont {Takeuchi}}, \bibinfo {author} {\bibfnamefont
  {T.}~\bibnamefont {Takahashi}}, \bibinfo {author} {\bibfnamefont
  {T.}~\bibnamefont {Mochiku}}, \bibinfo {author} {\bibfnamefont
  {K.}~\bibnamefont {Kadowaki}}, \bibinfo {author} {\bibfnamefont
  {P.}~\bibnamefont {Guptasarma}},\ and\ \bibinfo {author} {\bibfnamefont
  {D.~G.}\ \bibnamefont {Hinks}},\ }\href {https://doi.org/10.1038/32366}
  {\bibfield  {journal} {\bibinfo  {journal} {Nature}\ }\textbf {\bibinfo
  {volume} {392}},\ \bibinfo {pages} {157} (\bibinfo {year}
  {1998})}\BibitemShut {NoStop}%
\bibitem [{\citenamefont {Doiron-Leyraud}\ \emph {et~al.}(2007)\citenamefont
  {Doiron-Leyraud}, \citenamefont {Proust}, \citenamefont {LeBoeuf},
  \citenamefont {Levallois}, \citenamefont {Bonnemaison}, \citenamefont
  {Liang}, \citenamefont {Bonn}, \citenamefont {Hardy},\ and\ \citenamefont
  {Taillefer}}]{DoironLeyraud2007}%
  \BibitemOpen
  \bibfield  {author} {\bibinfo {author} {\bibfnamefont {N.}~\bibnamefont
  {Doiron-Leyraud}}, \bibinfo {author} {\bibfnamefont {C.}~\bibnamefont
  {Proust}}, \bibinfo {author} {\bibfnamefont {D.}~\bibnamefont {LeBoeuf}},
  \bibinfo {author} {\bibfnamefont {J.}~\bibnamefont {Levallois}}, \bibinfo
  {author} {\bibfnamefont {J.-B.}\ \bibnamefont {Bonnemaison}}, \bibinfo
  {author} {\bibfnamefont {R.}~\bibnamefont {Liang}}, \bibinfo {author}
  {\bibfnamefont {D.~A.}\ \bibnamefont {Bonn}}, \bibinfo {author}
  {\bibfnamefont {W.~N.}\ \bibnamefont {Hardy}},\ and\ \bibinfo {author}
  {\bibfnamefont {L.}~\bibnamefont {Taillefer}},\ }\href
  {https://doi.org/10.1038/nature05872} {\bibfield  {journal} {\bibinfo
  {journal} {Nature}\ }\textbf {\bibinfo {volume} {447}},\ \bibinfo {pages}
  {565} (\bibinfo {year} {2007})}\BibitemShut {NoStop}%
\bibitem [{\citenamefont {Badoux}\ \emph {et~al.}(2016)\citenamefont {Badoux},
  \citenamefont {Tabis}, \citenamefont {Lalibert{\'e}}, \citenamefont
  {Grissonnanche}, \citenamefont {Vignolle}, \citenamefont {Vignolles},
  \citenamefont {B{\'e}ard}, \citenamefont {Bonn}, \citenamefont {Hardy},
  \citenamefont {Liang}, \citenamefont {Doiron-Leyraud}, \citenamefont
  {Taillefer},\ and\ \citenamefont {Proust}}]{Badoux2016}%
  \BibitemOpen
  \bibfield  {author} {\bibinfo {author} {\bibfnamefont {S.}~\bibnamefont
  {Badoux}}, \bibinfo {author} {\bibfnamefont {W.}~\bibnamefont {Tabis}},
  \bibinfo {author} {\bibfnamefont {F.}~\bibnamefont {Lalibert{\'e}}}, \bibinfo
  {author} {\bibfnamefont {G.}~\bibnamefont {Grissonnanche}}, \bibinfo {author}
  {\bibfnamefont {B.}~\bibnamefont {Vignolle}}, \bibinfo {author}
  {\bibfnamefont {D.}~\bibnamefont {Vignolles}}, \bibinfo {author}
  {\bibfnamefont {J.}~\bibnamefont {B{\'e}ard}}, \bibinfo {author}
  {\bibfnamefont {D.~A.}\ \bibnamefont {Bonn}}, \bibinfo {author}
  {\bibfnamefont {W.~N.}\ \bibnamefont {Hardy}}, \bibinfo {author}
  {\bibfnamefont {R.}~\bibnamefont {Liang}}, \bibinfo {author} {\bibfnamefont
  {N.}~\bibnamefont {Doiron-Leyraud}}, \bibinfo {author} {\bibfnamefont
  {L.}~\bibnamefont {Taillefer}},\ and\ \bibinfo {author} {\bibfnamefont
  {C.}~\bibnamefont {Proust}},\ }\href {https://doi.org/10.1038/nature16983}
  {\bibfield  {journal} {\bibinfo  {journal} {Nature}\ }\textbf {\bibinfo
  {volume} {531}},\ \bibinfo {pages} {210} (\bibinfo {year}
  {2016})}\BibitemShut {NoStop}%
\bibitem [{\citenamefont {Keimer}\ \emph {et~al.}(2015)\citenamefont {Keimer},
  \citenamefont {Kivelson}, \citenamefont {Norman}, \citenamefont {Uchida},\
  and\ \citenamefont {Zaanen}}]{Keimer2015}%
  \BibitemOpen
  \bibfield  {author} {\bibinfo {author} {\bibfnamefont {B.}~\bibnamefont
  {Keimer}}, \bibinfo {author} {\bibfnamefont {S.~A.}\ \bibnamefont
  {Kivelson}}, \bibinfo {author} {\bibfnamefont {M.~R.}\ \bibnamefont
  {Norman}}, \bibinfo {author} {\bibfnamefont {S.}~\bibnamefont {Uchida}},\
  and\ \bibinfo {author} {\bibfnamefont {J.}~\bibnamefont {Zaanen}},\ }\href
  {https://doi.org/10.1038/nature14165} {\bibfield  {journal} {\bibinfo
  {journal} {Nature}\ }\textbf {\bibinfo {volume} {518}},\ \bibinfo {pages}
  {179} (\bibinfo {year} {2015})}\BibitemShut {NoStop}%
\bibitem [{\citenamefont {Kokkinis}\ and\ \citenamefont
  {Chubukov}(2025)}]{Andrei-NatCom2025}%
  \BibitemOpen
  \bibfield  {author} {\bibinfo {author} {\bibfnamefont {E.~K.}\ \bibnamefont
  {Kokkinis}}\ and\ \bibinfo {author} {\bibfnamefont {A.~V.}\ \bibnamefont
  {Chubukov}},\ }\href {https://doi.org/10.1038/s41467-025-67835-w} {\bibfield
  {journal} {\bibinfo  {journal} {Nat. Commun.}\ }\textbf {\bibinfo {volume}
  {17}},\ \bibinfo {pages} {1075} (\bibinfo {year} {2025})}\BibitemShut
  {NoStop}%
\bibitem [{\citenamefont {Fang}\ \emph {et~al.}(2022)\citenamefont {Fang},
  \citenamefont {Grissonnanche}, \citenamefont {Legros}, \citenamefont
  {Verret}, \citenamefont {Lalibert{\'e}}, \citenamefont {Collignon},
  \citenamefont {Ataei}, \citenamefont {Dion}, \citenamefont {Zhou},
  \citenamefont {Graf}, \citenamefont {Lawler}, \citenamefont {Goddard},
  \citenamefont {Taillefer},\ and\ \citenamefont {Ramshaw}}]{Fang-NatPhys2022}%
  \BibitemOpen
  \bibfield  {author} {\bibinfo {author} {\bibfnamefont {Y.}~\bibnamefont
  {Fang}}, \bibinfo {author} {\bibfnamefont {G.}~\bibnamefont {Grissonnanche}},
  \bibinfo {author} {\bibfnamefont {A.}~\bibnamefont {Legros}}, \bibinfo
  {author} {\bibfnamefont {S.}~\bibnamefont {Verret}}, \bibinfo {author}
  {\bibfnamefont {F.}~\bibnamefont {Lalibert{\'e}}}, \bibinfo {author}
  {\bibfnamefont {C.}~\bibnamefont {Collignon}}, \bibinfo {author}
  {\bibfnamefont {A.}~\bibnamefont {Ataei}}, \bibinfo {author} {\bibfnamefont
  {M.}~\bibnamefont {Dion}}, \bibinfo {author} {\bibfnamefont {J.}~\bibnamefont
  {Zhou}}, \bibinfo {author} {\bibfnamefont {D.}~\bibnamefont {Graf}}, \bibinfo
  {author} {\bibfnamefont {M.~J.}\ \bibnamefont {Lawler}}, \bibinfo {author}
  {\bibfnamefont {P.~A.}\ \bibnamefont {Goddard}}, \bibinfo {author}
  {\bibfnamefont {L.}~\bibnamefont {Taillefer}},\ and\ \bibinfo {author}
  {\bibfnamefont {B.~J.}\ \bibnamefont {Ramshaw}},\ }\href
  {https://doi.org/10.1038/s41567-022-01514-1} {\bibfield  {journal} {\bibinfo
  {journal} {Nat. Phys.}\ }\textbf {\bibinfo {volume} {18}},\ \bibinfo {pages}
  {558} (\bibinfo {year} {2022})}\BibitemShut {NoStop}%
\bibitem [{\citenamefont {Chan}\ \emph {et~al.}(2025)\citenamefont {Chan},
  \citenamefont {Schreiber}, \citenamefont {Ayala-Valenzuela}, \citenamefont
  {Bauer}, \citenamefont {Shekhter},\ and\ \citenamefont
  {Harrison}}]{Chan-NatPhys2025}%
  \BibitemOpen
  \bibfield  {author} {\bibinfo {author} {\bibfnamefont {M.~K.}\ \bibnamefont
  {Chan}}, \bibinfo {author} {\bibfnamefont {K.~A.}\ \bibnamefont {Schreiber}},
  \bibinfo {author} {\bibfnamefont {O.~E.}\ \bibnamefont {Ayala-Valenzuela}},
  \bibinfo {author} {\bibfnamefont {E.~D.}\ \bibnamefont {Bauer}}, \bibinfo
  {author} {\bibfnamefont {A.}~\bibnamefont {Shekhter}},\ and\ \bibinfo
  {author} {\bibfnamefont {N.}~\bibnamefont {Harrison}},\ }\href
  {https://doi.org/10.1038/s41567-025-03032-2} {\bibfield  {journal} {\bibinfo
  {journal} {Nat. Phys.}\ }\textbf {\bibinfo {volume} {21}},\ \bibinfo {pages}
  {1753} (\bibinfo {year} {2025})}\BibitemShut {NoStop}%
\bibitem [{\citenamefont {Luttinger}(1960)}]{Luttinger}%
  \BibitemOpen
  \bibfield  {author} {\bibinfo {author} {\bibfnamefont {J.~M.}\ \bibnamefont
  {Luttinger}},\ }\href {https://doi.org/10.1103/PhysRev.119.1153} {\bibfield
  {journal} {\bibinfo  {journal} {Phys. Rev.}\ }\textbf {\bibinfo {volume}
  {119}},\ \bibinfo {pages} {1153} (\bibinfo {year} {1960})}\BibitemShut
  {NoStop}%
\bibitem [{\citenamefont {Yang}\ \emph {et~al.}(2006)\citenamefont {Yang},
  \citenamefont {Rice},\ and\ \citenamefont {Zhang}}]{YangRiceZhang2006}%
  \BibitemOpen
  \bibfield  {author} {\bibinfo {author} {\bibfnamefont {K.-Y.}\ \bibnamefont
  {Yang}}, \bibinfo {author} {\bibfnamefont {T.~M.}\ \bibnamefont {Rice}},\
  and\ \bibinfo {author} {\bibfnamefont {F.-C.}\ \bibnamefont {Zhang}},\ }\href
  {https://doi.org/10.1103/PhysRevB.73.174501} {\bibfield  {journal} {\bibinfo
  {journal} {Phys. Rev. B}\ }\textbf {\bibinfo {volume} {73}},\ \bibinfo
  {pages} {174501} (\bibinfo {year} {2006})}\BibitemShut {NoStop}%
\bibitem [{\citenamefont {Heath}\ and\ \citenamefont
  {Bedell}(2020)}]{Heath_2020}%
  \BibitemOpen
  \bibfield  {author} {\bibinfo {author} {\bibfnamefont {J.~T.}\ \bibnamefont
  {Heath}}\ and\ \bibinfo {author} {\bibfnamefont {K.~S.}\ \bibnamefont
  {Bedell}},\ }\href {https://doi.org/10.1088/1367-2630/ab890e} {\bibfield
  {journal} {\bibinfo  {journal} {New J. Phys.}\ }\textbf {\bibinfo {volume}
  {22}},\ \bibinfo {pages} {063011} (\bibinfo {year} {2020})}\BibitemShut
  {NoStop}%
\bibitem [{\citenamefont {Skolimowski}\ and\ \citenamefont
  {Fabrizio}(2022)}]{Jan-PRB2022}%
  \BibitemOpen
  \bibfield  {author} {\bibinfo {author} {\bibfnamefont {J.}~\bibnamefont
  {Skolimowski}}\ and\ \bibinfo {author} {\bibfnamefont {M.}~\bibnamefont
  {Fabrizio}},\ }\href {https://doi.org/10.1103/PhysRevB.106.045109} {\bibfield
   {journal} {\bibinfo  {journal} {Phys. Rev. B}\ }\textbf {\bibinfo {volume}
  {106}},\ \bibinfo {pages} {045109} (\bibinfo {year} {2022})}\BibitemShut
  {NoStop}%
\bibitem [{\citenamefont {Fabrizio}(2022)}]{mio-2}%
  \BibitemOpen
  \bibfield  {author} {\bibinfo {author} {\bibfnamefont {M.}~\bibnamefont
  {Fabrizio}},\ }\href {https://doi.org/10.1038/s41467-022-29190-y} {\bibfield
  {journal} {\bibinfo  {journal} {Nat. Commun.}\ }\textbf {\bibinfo {volume}
  {13}},\ \bibinfo {pages} {1561} (\bibinfo {year} {2022})}\BibitemShut
  {NoStop}%
\bibitem [{\citenamefont {Anderson}(1987)}]{Anderson1987}%
  \BibitemOpen
  \bibfield  {author} {\bibinfo {author} {\bibfnamefont {P.~W.}\ \bibnamefont
  {Anderson}},\ }\href {https://doi.org/10.1126/science.235.4793.1196}
  {\bibfield  {journal} {\bibinfo  {journal} {Science}\ }\textbf {\bibinfo
  {volume} {235}},\ \bibinfo {pages} {1196} (\bibinfo {year}
  {1987})}\BibitemShut {NoStop}%
\bibitem [{\citenamefont {Baskaran}\ \emph {et~al.}(1987)\citenamefont
  {Baskaran}, \citenamefont {Zou},\ and\ \citenamefont
  {Anderson}}]{BaskaranZouAnderson1987}%
  \BibitemOpen
  \bibfield  {author} {\bibinfo {author} {\bibfnamefont {G.}~\bibnamefont
  {Baskaran}}, \bibinfo {author} {\bibfnamefont {Z.}~\bibnamefont {Zou}},\ and\
  \bibinfo {author} {\bibfnamefont {P.~W.}\ \bibnamefont {Anderson}},\ }\href
  {https://doi.org/10.1016/0038-1098(87)90642-9} {\bibfield  {journal}
  {\bibinfo  {journal} {Solid State Commun.}\ }\textbf {\bibinfo {volume}
  {63}},\ \bibinfo {pages} {973} (\bibinfo {year} {1987})}\BibitemShut
  {NoStop}%
\bibitem [{\citenamefont {Lee}\ \emph {et~al.}(2006)\citenamefont {Lee},
  \citenamefont {Nagaosa},\ and\ \citenamefont {Wen}}]{LeeNagaosaWen2006}%
  \BibitemOpen
  \bibfield  {author} {\bibinfo {author} {\bibfnamefont {P.~A.}\ \bibnamefont
  {Lee}}, \bibinfo {author} {\bibfnamefont {N.}~\bibnamefont {Nagaosa}},\ and\
  \bibinfo {author} {\bibfnamefont {X.-G.}\ \bibnamefont {Wen}},\ }\href
  {https://doi.org/10.1103/RevModPhys.78.17} {\bibfield  {journal} {\bibinfo
  {journal} {Rev. Mod. Phys.}\ }\textbf {\bibinfo {volume} {78}},\ \bibinfo
  {pages} {17} (\bibinfo {year} {2006})}\BibitemShut {NoStop}%
\bibitem [{\citenamefont {Kotliar}\ and\ \citenamefont
  {Liu}(1988)}]{LiuKotliar1988}%
  \BibitemOpen
  \bibfield  {author} {\bibinfo {author} {\bibfnamefont {G.}~\bibnamefont
  {Kotliar}}\ and\ \bibinfo {author} {\bibfnamefont {J.}~\bibnamefont {Liu}},\
  }\href {https://doi.org/10.1103/PhysRevB.38.5142} {\bibfield  {journal}
  {\bibinfo  {journal} {Phys. Rev. B}\ }\textbf {\bibinfo {volume} {38}},\
  \bibinfo {pages} {5142} (\bibinfo {year} {1988})}\BibitemShut {NoStop}%
\bibitem [{\citenamefont {Ribeiro}\ and\ \citenamefont
  {Wen}(2005)}]{Wen-PRL2005}%
  \BibitemOpen
  \bibfield  {author} {\bibinfo {author} {\bibfnamefont {T.~C.}\ \bibnamefont
  {Ribeiro}}\ and\ \bibinfo {author} {\bibfnamefont {X.-G.}\ \bibnamefont
  {Wen}},\ }\href {https://doi.org/10.1103/PhysRevLett.95.057001} {\bibfield
  {journal} {\bibinfo  {journal} {Phys. Rev. Lett.}\ }\textbf {\bibinfo
  {volume} {95}},\ \bibinfo {pages} {057001} (\bibinfo {year}
  {2005})}\BibitemShut {NoStop}%
\bibitem [{\citenamefont {Ng}(2005)}]{Ng-PRB2005}%
  \BibitemOpen
  \bibfield  {author} {\bibinfo {author} {\bibfnamefont {T.-K.}\ \bibnamefont
  {Ng}},\ }\href {https://doi.org/10.1103/PhysRevB.71.172509} {\bibfield
  {journal} {\bibinfo  {journal} {Phys. Rev. B}\ }\textbf {\bibinfo {volume}
  {71}},\ \bibinfo {pages} {172509} (\bibinfo {year} {2005})}\BibitemShut
  {NoStop}%
\bibitem [{\citenamefont {Punk}\ \emph {et~al.}(2015)\citenamefont {Punk},
  \citenamefont {Allais},\ and\ \citenamefont {Sachdev}}]{Punk-PNAS2015}%
  \BibitemOpen
  \bibfield  {author} {\bibinfo {author} {\bibfnamefont {M.}~\bibnamefont
  {Punk}}, \bibinfo {author} {\bibfnamefont {A.}~\bibnamefont {Allais}},\ and\
  \bibinfo {author} {\bibfnamefont {S.}~\bibnamefont {Sachdev}},\ }\href
  {https://doi.org/10.1073/pnas.1512206112} {\bibfield  {journal} {\bibinfo
  {journal} {Proc. Natl. Acad. Sci. USA}\ }\textbf {\bibinfo {volume} {112}},\
  \bibinfo {pages} {9552} (\bibinfo {year} {2015})}\BibitemShut {NoStop}%
\bibitem [{\citenamefont {Scheurer}\ \emph {et~al.}(2018)\citenamefont
  {Scheurer}, \citenamefont {Chatterjee}, \citenamefont {Wu}, \citenamefont
  {Ferrero}, \citenamefont {Georges},\ and\ \citenamefont
  {Sachdev}}]{Georges-PNAS2018}%
  \BibitemOpen
  \bibfield  {author} {\bibinfo {author} {\bibfnamefont {M.~S.}\ \bibnamefont
  {Scheurer}}, \bibinfo {author} {\bibfnamefont {S.}~\bibnamefont
  {Chatterjee}}, \bibinfo {author} {\bibfnamefont {W.}~\bibnamefont {Wu}},
  \bibinfo {author} {\bibfnamefont {M.}~\bibnamefont {Ferrero}}, \bibinfo
  {author} {\bibfnamefont {A.}~\bibnamefont {Georges}},\ and\ \bibinfo {author}
  {\bibfnamefont {S.}~\bibnamefont {Sachdev}},\ }\href
  {https://doi.org/10.1073/pnas.1720580115} {\bibfield  {journal} {\bibinfo
  {journal} {Proc. Natl. Acad. Sci. USA}\ }\textbf {\bibinfo {volume} {115}},\
  \bibinfo {pages} {E3665} (\bibinfo {year} {2018})}\BibitemShut {NoStop}%
\bibitem [{\citenamefont {Wu}\ \emph {et~al.}(2018)\citenamefont {Wu},
  \citenamefont {Scheurer}, \citenamefont {Chatterjee}, \citenamefont
  {Sachdev}, \citenamefont {Georges},\ and\ \citenamefont
  {Ferrero}}]{Georges-PRX2018}%
  \BibitemOpen
  \bibfield  {author} {\bibinfo {author} {\bibfnamefont {W.}~\bibnamefont
  {Wu}}, \bibinfo {author} {\bibfnamefont {M.~S.}\ \bibnamefont {Scheurer}},
  \bibinfo {author} {\bibfnamefont {S.}~\bibnamefont {Chatterjee}}, \bibinfo
  {author} {\bibfnamefont {S.}~\bibnamefont {Sachdev}}, \bibinfo {author}
  {\bibfnamefont {A.}~\bibnamefont {Georges}},\ and\ \bibinfo {author}
  {\bibfnamefont {M.}~\bibnamefont {Ferrero}},\ }\href
  {https://doi.org/10.1103/PhysRevX.8.021048} {\bibfield  {journal} {\bibinfo
  {journal} {Phys. Rev. X}\ }\textbf {\bibinfo {volume} {8}},\ \bibinfo {pages}
  {021048} (\bibinfo {year} {2018})}\BibitemShut {NoStop}%
\bibitem [{\citenamefont {Brunkert}\ and\ \citenamefont
  {Punk}(2020)}]{Punk-PRR2020}%
  \BibitemOpen
  \bibfield  {author} {\bibinfo {author} {\bibfnamefont {J.}~\bibnamefont
  {Brunkert}}\ and\ \bibinfo {author} {\bibfnamefont {M.}~\bibnamefont
  {Punk}},\ }\href {https://doi.org/10.1103/PhysRevResearch.2.043019}
  {\bibfield  {journal} {\bibinfo  {journal} {Phys. Rev. Res.}\ }\textbf
  {\bibinfo {volume} {2}},\ \bibinfo {pages} {043019} (\bibinfo {year}
  {2020})}\BibitemShut {NoStop}%
\bibitem [{\citenamefont {Nyhegn}\ \emph {et~al.}(2025)\citenamefont {Nyhegn},
  \citenamefont {Nielsen}, \citenamefont {Balents},\ and\ \citenamefont
  {Bruun}}]{Nyhegn-PRXQ2025}%
  \BibitemOpen
  \bibfield  {author} {\bibinfo {author} {\bibfnamefont {J.~H.}\ \bibnamefont
  {Nyhegn}}, \bibinfo {author} {\bibfnamefont {K.~K.}\ \bibnamefont {Nielsen}},
  \bibinfo {author} {\bibfnamefont {L.}~\bibnamefont {Balents}},\ and\ \bibinfo
  {author} {\bibfnamefont {G.~M.}\ \bibnamefont {Bruun}},\ }\href
  {https://doi.org/10.1103/w23h-dhrk} {\bibfield  {journal} {\bibinfo
  {journal} {PRX Quantum}\ }\textbf {\bibinfo {volume} {6}},\ \bibinfo {pages}
  {040347} (\bibinfo {year} {2025})}\BibitemShut {NoStop}%
\bibitem [{\citenamefont {Senthil}\ \emph {et~al.}(2003)\citenamefont
  {Senthil}, \citenamefont {Sachdev},\ and\ \citenamefont
  {Vojta}}]{SenthilSachdevVojta2003}%
  \BibitemOpen
  \bibfield  {author} {\bibinfo {author} {\bibfnamefont {T.}~\bibnamefont
  {Senthil}}, \bibinfo {author} {\bibfnamefont {S.}~\bibnamefont {Sachdev}},\
  and\ \bibinfo {author} {\bibfnamefont {M.}~\bibnamefont {Vojta}},\ }\href
  {https://doi.org/10.1103/PhysRevLett.90.216403} {\bibfield  {journal}
  {\bibinfo  {journal} {Phys. Rev. Lett.}\ }\textbf {\bibinfo {volume} {90}},\
  \bibinfo {pages} {216403} (\bibinfo {year} {2003})}\BibitemShut {NoStop}%
\bibitem [{\citenamefont {Senthil}\ \emph {et~al.}(2004)\citenamefont
  {Senthil}, \citenamefont {Vojta},\ and\ \citenamefont
  {Sachdev}}]{SenthilVojtaSachdev2004}%
  \BibitemOpen
  \bibfield  {author} {\bibinfo {author} {\bibfnamefont {T.}~\bibnamefont
  {Senthil}}, \bibinfo {author} {\bibfnamefont {M.}~\bibnamefont {Vojta}},\
  and\ \bibinfo {author} {\bibfnamefont {S.}~\bibnamefont {Sachdev}},\ }\href
  {https://doi.org/10.1103/PhysRevB.69.035111} {\bibfield  {journal} {\bibinfo
  {journal} {Phys. Rev. B}\ }\textbf {\bibinfo {volume} {69}},\ \bibinfo
  {pages} {035111} (\bibinfo {year} {2004})}\BibitemShut {NoStop}%
\bibitem [{\citenamefont {Zhang}\ and\ \citenamefont
  {Sachdev}(2020)}]{ZhangSachdev2020}%
  \BibitemOpen
  \bibfield  {author} {\bibinfo {author} {\bibfnamefont {Y.-H.}\ \bibnamefont
  {Zhang}}\ and\ \bibinfo {author} {\bibfnamefont {S.}~\bibnamefont
  {Sachdev}},\ }\href {https://doi.org/10.1103/PhysRevResearch.2.023172}
  {\bibfield  {journal} {\bibinfo  {journal} {Phys. Rev. Res.}\ }\textbf
  {\bibinfo {volume} {2}},\ \bibinfo {pages} {023172} (\bibinfo {year}
  {2020})}\BibitemShut {NoStop}%
\bibitem [{\citenamefont {Zhou}\ \emph {et~al.}(2025)\citenamefont {Zhou},
  \citenamefont {Jin},\ and\ \citenamefont {Zhang}}]{Zhou-PRB2025}%
  \BibitemOpen
  \bibfield  {author} {\bibinfo {author} {\bibfnamefont {B.}~\bibnamefont
  {Zhou}}, \bibinfo {author} {\bibfnamefont {H.-K.}\ \bibnamefont {Jin}},\ and\
  \bibinfo {author} {\bibfnamefont {Y.-H.}\ \bibnamefont {Zhang}},\ }\href
  {https://doi.org/10.1103/8knn-mr5x} {\bibfield  {journal} {\bibinfo
  {journal} {Phys. Rev. B}\ }\textbf {\bibinfo {volume} {112}},\ \bibinfo
  {pages} {115159} (\bibinfo {year} {2025})}\BibitemShut {NoStop}%
\bibitem [{\citenamefont {Bonetti}\ \emph {et~al.}(2026)\citenamefont
  {Bonetti}, \citenamefont {Christos}, \citenamefont {Nikolaenko},
  \citenamefont {Patel},\ and\ \citenamefont {Sachdev}}]{Bonetti_2026}%
  \BibitemOpen
  \bibfield  {author} {\bibinfo {author} {\bibfnamefont {P.~M.}\ \bibnamefont
  {Bonetti}}, \bibinfo {author} {\bibfnamefont {M.}~\bibnamefont {Christos}},
  \bibinfo {author} {\bibfnamefont {A.}~\bibnamefont {Nikolaenko}}, \bibinfo
  {author} {\bibfnamefont {A.~A.}\ \bibnamefont {Patel}},\ and\ \bibinfo
  {author} {\bibfnamefont {S.}~\bibnamefont {Sachdev}},\ }\href
  {https://doi.org/10.1088/1361-6633/ae530d} {\bibfield  {journal} {\bibinfo
  {journal} {Rep. Prog. Phys.}\ }\textbf {\bibinfo {volume} {89}},\ \bibinfo
  {pages} {044501} (\bibinfo {year} {2026})}\BibitemShut {NoStop}%
\bibitem [{\citenamefont {Lanat\`a}\ \emph {et~al.}(2017)\citenamefont
  {Lanat\`a}, \citenamefont {Lee}, \citenamefont {Yao},\ and\ \citenamefont
  {Dobrosavljevi\ifmmode~\acute{c}\else \'{c}\fi{}}}]{Lanata2017}%
  \BibitemOpen
  \bibfield  {author} {\bibinfo {author} {\bibfnamefont {N.}~\bibnamefont
  {Lanat\`a}}, \bibinfo {author} {\bibfnamefont {T.-H.}\ \bibnamefont {Lee}},
  \bibinfo {author} {\bibfnamefont {Y.-X.}\ \bibnamefont {Yao}},\ and\ \bibinfo
  {author} {\bibfnamefont {V.}~\bibnamefont
  {Dobrosavljevi\ifmmode~\acute{c}\else \'{c}\fi{}}},\ }\href
  {https://doi.org/10.1103/PhysRevB.96.195126} {\bibfield  {journal} {\bibinfo
  {journal} {Phys. Rev. B}\ }\textbf {\bibinfo {volume} {96}},\ \bibinfo
  {pages} {195126} (\bibinfo {year} {2017})}\BibitemShut {NoStop}%
\bibitem [{\citenamefont {Guerci}\ \emph {et~al.}(2019)\citenamefont {Guerci},
  \citenamefont {Capone},\ and\ \citenamefont {Fabrizio}}]{guerci2019}%
  \BibitemOpen
  \bibfield  {author} {\bibinfo {author} {\bibfnamefont {D.}~\bibnamefont
  {Guerci}}, \bibinfo {author} {\bibfnamefont {M.}~\bibnamefont {Capone}},\
  and\ \bibinfo {author} {\bibfnamefont {M.}~\bibnamefont {Fabrizio}},\ }\href
  {https://doi.org/10.1103/PhysRevMaterials.3.054605} {\bibfield  {journal}
  {\bibinfo  {journal} {Phys. Rev. Mater.}\ }\textbf {\bibinfo {volume} {3}},\
  \bibinfo {pages} {054605} (\bibinfo {year} {2019})}\BibitemShut {NoStop}%
\bibitem [{\citenamefont {Frank}\ \emph {et~al.}(2021)\citenamefont {Frank},
  \citenamefont {Lee}, \citenamefont {Bhattacharyya}, \citenamefont {Tsang},
  \citenamefont {Quito}, \citenamefont {Dobrosavljevi\ifmmode~\acute{c}\else
  \'{c}\fi{}}, \citenamefont {Christiansen},\ and\ \citenamefont
  {Lanat\`a}}]{frank2021}%
  \BibitemOpen
  \bibfield  {author} {\bibinfo {author} {\bibfnamefont {M.~S.}\ \bibnamefont
  {Frank}}, \bibinfo {author} {\bibfnamefont {T.-H.}\ \bibnamefont {Lee}},
  \bibinfo {author} {\bibfnamefont {G.}~\bibnamefont {Bhattacharyya}}, \bibinfo
  {author} {\bibfnamefont {P.~K.~H.}\ \bibnamefont {Tsang}}, \bibinfo {author}
  {\bibfnamefont {V.~L.}\ \bibnamefont {Quito}}, \bibinfo {author}
  {\bibfnamefont {V.}~\bibnamefont {Dobrosavljevi\ifmmode~\acute{c}\else
  \'{c}\fi{}}}, \bibinfo {author} {\bibfnamefont {O.}~\bibnamefont
  {Christiansen}},\ and\ \bibinfo {author} {\bibfnamefont {N.}~\bibnamefont
  {Lanat\`a}},\ }\href {https://doi.org/10.1103/PhysRevB.104.L081103}
  {\bibfield  {journal} {\bibinfo  {journal} {Phys. Rev. B}\ }\textbf {\bibinfo
  {volume} {104}},\ \bibinfo {pages} {L081103} (\bibinfo {year}
  {2021})}\BibitemShut {NoStop}%
\bibitem [{\citenamefont {Lee}\ \emph {et~al.}(2023{\natexlab{a}})\citenamefont
  {Lee}, \citenamefont {Melnick}, \citenamefont {Adler}, \citenamefont
  {Lanat\`a},\ and\ \citenamefont {Kotliar}}]{Lee2023a}%
  \BibitemOpen
  \bibfield  {author} {\bibinfo {author} {\bibfnamefont {T.-H.}\ \bibnamefont
  {Lee}}, \bibinfo {author} {\bibfnamefont {C.}~\bibnamefont {Melnick}},
  \bibinfo {author} {\bibfnamefont {R.}~\bibnamefont {Adler}}, \bibinfo
  {author} {\bibfnamefont {N.}~\bibnamefont {Lanat\`a}},\ and\ \bibinfo
  {author} {\bibfnamefont {G.}~\bibnamefont {Kotliar}},\ }\href
  {https://doi.org/10.1103/PhysRevB.108.245147} {\bibfield  {journal} {\bibinfo
   {journal} {Phys. Rev. B}\ }\textbf {\bibinfo {volume} {108}},\ \bibinfo
  {pages} {245147} (\bibinfo {year} {2023}{\natexlab{a}})}\BibitemShut
  {NoStop}%
\bibitem [{\citenamefont {Lee}\ \emph {et~al.}(2023{\natexlab{b}})\citenamefont
  {Lee}, \citenamefont {Lanat\`a},\ and\ \citenamefont {Kotliar}}]{Lee2023b}%
  \BibitemOpen
  \bibfield  {author} {\bibinfo {author} {\bibfnamefont {T.-H.}\ \bibnamefont
  {Lee}}, \bibinfo {author} {\bibfnamefont {N.}~\bibnamefont {Lanat\`a}},\ and\
  \bibinfo {author} {\bibfnamefont {G.}~\bibnamefont {Kotliar}},\ }\href
  {https://doi.org/10.1103/PhysRevB.107.L121104} {\bibfield  {journal}
  {\bibinfo  {journal} {Phys. Rev. B}\ }\textbf {\bibinfo {volume} {107}},\
  \bibinfo {pages} {L121104} (\bibinfo {year}
  {2023}{\natexlab{b}})}\BibitemShut {NoStop}%
\bibitem [{\citenamefont {Mejuto-Zaera}\ and\ \citenamefont
  {Fabrizio}(2023)}]{Mejuto2023a}%
  \BibitemOpen
  \bibfield  {author} {\bibinfo {author} {\bibfnamefont {C.}~\bibnamefont
  {Mejuto-Zaera}}\ and\ \bibinfo {author} {\bibfnamefont {M.}~\bibnamefont
  {Fabrizio}},\ }\href {https://doi.org/10.1103/PhysRevB.107.235150} {\bibfield
   {journal} {\bibinfo  {journal} {Phys. Rev. B}\ }\textbf {\bibinfo {volume}
  {107}},\ \bibinfo {pages} {235150} (\bibinfo {year} {2023})}\BibitemShut
  {NoStop}%
\bibitem [{\citenamefont {Tagliente}\ \emph {et~al.}(2025)\citenamefont
  {Tagliente}, \citenamefont {Mejuto-Zaera},\ and\ \citenamefont
  {Fabrizio}}]{Tagliente2025}%
  \BibitemOpen
  \bibfield  {author} {\bibinfo {author} {\bibfnamefont {A.~M.}\ \bibnamefont
  {Tagliente}}, \bibinfo {author} {\bibfnamefont {C.}~\bibnamefont
  {Mejuto-Zaera}},\ and\ \bibinfo {author} {\bibfnamefont {M.}~\bibnamefont
  {Fabrizio}},\ }\href {https://doi.org/10.1103/PhysRevB.111.125110} {\bibfield
   {journal} {\bibinfo  {journal} {Phys. Rev. B}\ }\textbf {\bibinfo {volume}
  {111}},\ \bibinfo {pages} {125110} (\bibinfo {year} {2025})}\BibitemShut
  {NoStop}%
\bibitem [{\citenamefont {Pasqua}\ \emph {et~al.}(2026)\citenamefont {Pasqua},
  \citenamefont {Tagliente}, \citenamefont {Bellomia}, \citenamefont
  {Monserrat}, \citenamefont {Fabrizio},\ and\ \citenamefont
  {Mejuto-Zaera}}]{Pasqua2026}%
  \BibitemOpen
  \bibfield  {author} {\bibinfo {author} {\bibfnamefont {I.}~\bibnamefont
  {Pasqua}}, \bibinfo {author} {\bibfnamefont {A.~M.}\ \bibnamefont
  {Tagliente}}, \bibinfo {author} {\bibfnamefont {G.}~\bibnamefont {Bellomia}},
  \bibinfo {author} {\bibfnamefont {B.}~\bibnamefont {Monserrat}}, \bibinfo
  {author} {\bibfnamefont {M.}~\bibnamefont {Fabrizio}},\ and\ \bibinfo
  {author} {\bibfnamefont {C.}~\bibnamefont {Mejuto-Zaera}},\ }\bibfield
  {journal} {\bibinfo  {journal} {Phys. Rev. B}\ }\textbf {\bibinfo {volume}
  {113}},\ \href {https://doi.org/10.1103/dw49-y9vl} {10.1103/dw49-y9vl}
  (\bibinfo {year} {2026})\BibitemShut {NoStop}%
\bibitem [{\citenamefont {Mejuto-Zaera}\ and\ \citenamefont
  {Fabrizio}(2026)}]{Mejuto2026}%
  \BibitemOpen
  \bibfield  {author} {\bibinfo {author} {\bibfnamefont {C.}~\bibnamefont
  {Mejuto-Zaera}}\ and\ \bibinfo {author} {\bibfnamefont {M.}~\bibnamefont
  {Fabrizio}},\ }\href {https://doi.org/10.1063/5.0329844} {\bibfield
  {journal} {\bibinfo  {journal} {J. Chem. Phys.}\ }\textbf {\bibinfo {volume}
  {164}},\ \bibinfo {pages} {224125} (\bibinfo {year} {2026})}\BibitemShut
  {NoStop}%
\bibitem [{\citenamefont {Giuli}\ \emph
  {et~al.}(2025{\natexlab{a}})\citenamefont {Giuli}, \citenamefont {Hasan},
  \citenamefont {Kloss}, \citenamefont {Frank}, \citenamefont {Lee},
  \citenamefont {Gingras}, \citenamefont {Yao},\ and\ \citenamefont
  {Lanat\`a}}]{ArXivSamuele2026a}%
  \BibitemOpen
  \bibfield  {author} {\bibinfo {author} {\bibfnamefont {S.}~\bibnamefont
  {Giuli}}, \bibinfo {author} {\bibfnamefont {H.}~\bibnamefont {Hasan}},
  \bibinfo {author} {\bibfnamefont {B.}~\bibnamefont {Kloss}}, \bibinfo
  {author} {\bibfnamefont {M.~S.}\ \bibnamefont {Frank}}, \bibinfo {author}
  {\bibfnamefont {T.-H.}\ \bibnamefont {Lee}}, \bibinfo {author} {\bibfnamefont
  {O.}~\bibnamefont {Gingras}}, \bibinfo {author} {\bibfnamefont {Y.-X.}\
  \bibnamefont {Yao}},\ and\ \bibinfo {author} {\bibfnamefont {N.}~\bibnamefont
  {Lanat\`a}},\ }\href {https://doi.org/10.48550/ARXIV.2512.21666} {\bibinfo
  {title} {Linear foundation model for quantum embedding: Data-driven
  compression of the ghost gutzwiller variational space}} (\bibinfo {year}
  {2025}{\natexlab{a}})\BibitemShut {NoStop}%
\bibitem [{\citenamefont {Giuli}\ \emph {et~al.}(2026)\citenamefont {Giuli},
  \citenamefont {Lee}, \citenamefont {Yao}, \citenamefont {Kotliar},
  \citenamefont {Ruckenstein}, \citenamefont {Gingras},\ and\ \citenamefont
  {Lanat\`a}}]{ArXivSamuele2026b}%
  \BibitemOpen
  \bibfield  {author} {\bibinfo {author} {\bibfnamefont {S.}~\bibnamefont
  {Giuli}}, \bibinfo {author} {\bibfnamefont {T.-H.}\ \bibnamefont {Lee}},
  \bibinfo {author} {\bibfnamefont {Y.-X.}\ \bibnamefont {Yao}}, \bibinfo
  {author} {\bibfnamefont {G.}~\bibnamefont {Kotliar}}, \bibinfo {author}
  {\bibfnamefont {A.~E.}\ \bibnamefont {Ruckenstein}}, \bibinfo {author}
  {\bibfnamefont {O.}~\bibnamefont {Gingras}},\ and\ \bibinfo {author}
  {\bibfnamefont {N.}~\bibnamefont {Lanat\`a}},\ }\href
  {https://doi.org/10.48550/ARXIV.2603.20559} {\bibinfo {title} {Unifying
  variational and dynamical quantum embedding: From ghost gutzwiller
  approximation to dynamical mean-field theory}} (\bibinfo {year}
  {2026})\BibitemShut {NoStop}%
\bibitem [{\citenamefont {Kazemi-Moridani}\ \emph {et~al.}(2026)\citenamefont
  {Kazemi-Moridani}, \citenamefont {Giuli}, \citenamefont {Lee}, \citenamefont
  {Tremblay}, \citenamefont {C\^ot\'e}, \citenamefont {Lanat\`a},\ and\
  \citenamefont {Gingras}}]{ArXivMoridani2026}%
  \BibitemOpen
  \bibfield  {author} {\bibinfo {author} {\bibfnamefont {A.}~\bibnamefont
  {Kazemi-Moridani}}, \bibinfo {author} {\bibfnamefont {S.}~\bibnamefont
  {Giuli}}, \bibinfo {author} {\bibfnamefont {T.-H.}\ \bibnamefont {Lee}},
  \bibinfo {author} {\bibfnamefont {A.-M.~S.}\ \bibnamefont {Tremblay}},
  \bibinfo {author} {\bibfnamefont {M.}~\bibnamefont {C\^ot\'e}}, \bibinfo
  {author} {\bibfnamefont {N.}~\bibnamefont {Lanat\`a}},\ and\ \bibinfo
  {author} {\bibfnamefont {O.}~\bibnamefont {Gingras}},\ }\href
  {https://doi.org/10.48550/ARXIV.2604.24848} {\bibinfo {title} {Magnetic
  phases of the anisotropic triangular hubbard model from the ghost-gutzwiller
  approximation in the rotating spin-frame}} (\bibinfo {year}
  {2026})\BibitemShut {NoStop}%
\bibitem [{\citenamefont {Zhang}\ and\ \citenamefont
  {Rice}(1988)}]{ZhangRice1988}%
  \BibitemOpen
  \bibfield  {author} {\bibinfo {author} {\bibfnamefont {F.~C.}\ \bibnamefont
  {Zhang}}\ and\ \bibinfo {author} {\bibfnamefont {T.~M.}\ \bibnamefont
  {Rice}},\ }\href {https://doi.org/10.1103/PhysRevB.37.3759} {\bibfield
  {journal} {\bibinfo  {journal} {Phys. Rev. B}\ }\textbf {\bibinfo {volume}
  {37}},\ \bibinfo {pages} {3759} (\bibinfo {year} {1988})}\BibitemShut
  {NoStop}%
\bibitem [{\citenamefont {Gutzwiller}(1963)}]{Gutzwiller1963}%
  \BibitemOpen
  \bibfield  {author} {\bibinfo {author} {\bibfnamefont {M.~C.}\ \bibnamefont
  {Gutzwiller}},\ }\href {https://doi.org/10.1103/PhysRevLett.10.159}
  {\bibfield  {journal} {\bibinfo  {journal} {Phys. Rev. Lett.}\ }\textbf
  {\bibinfo {volume} {10}},\ \bibinfo {pages} {159} (\bibinfo {year}
  {1963})}\BibitemShut {NoStop}%
\bibitem [{\citenamefont {Gutzwiller}(1965)}]{Gutzwiller1965}%
  \BibitemOpen
  \bibfield  {author} {\bibinfo {author} {\bibfnamefont {M.~C.}\ \bibnamefont
  {Gutzwiller}},\ }\href {https://doi.org/10.1103/PhysRev.137.A1726} {\bibfield
   {journal} {\bibinfo  {journal} {Phys. Rev.}\ }\textbf {\bibinfo {volume}
  {137}},\ \bibinfo {pages} {A1726} (\bibinfo {year} {1965})}\BibitemShut
  {NoStop}%
\bibitem [{\citenamefont {B\"unemann}\ \emph {et~al.}(1998)\citenamefont
  {B\"unemann}, \citenamefont {Weber},\ and\ \citenamefont
  {Gebhard}}]{Bunemann1998}%
  \BibitemOpen
  \bibfield  {author} {\bibinfo {author} {\bibfnamefont {J.}~\bibnamefont
  {B\"unemann}}, \bibinfo {author} {\bibfnamefont {W.}~\bibnamefont {Weber}},\
  and\ \bibinfo {author} {\bibfnamefont {F.}~\bibnamefont {Gebhard}},\ }\href
  {https://doi.org/10.1103/PhysRevB.57.6896} {\bibfield  {journal} {\bibinfo
  {journal} {Phys. Rev. B}\ }\textbf {\bibinfo {volume} {57}},\ \bibinfo
  {pages} {6896} (\bibinfo {year} {1998})}\BibitemShut {NoStop}%
\bibitem [{\citenamefont {Fabrizio}(2007)}]{Fabrizio2007}%
  \BibitemOpen
  \bibfield  {author} {\bibinfo {author} {\bibfnamefont {M.}~\bibnamefont
  {Fabrizio}},\ }\href {https://doi.org/10.1103/PhysRevB.76.165110} {\bibfield
  {journal} {\bibinfo  {journal} {Phys. Rev. B}\ }\textbf {\bibinfo {volume}
  {76}},\ \bibinfo {pages} {165110} (\bibinfo {year} {2007})}\BibitemShut
  {NoStop}%
\bibitem [{\citenamefont {Yao}\ \emph {et~al.}(2014)\citenamefont {Yao},
  \citenamefont {Liu}, \citenamefont {Wang},\ and\ \citenamefont
  {Ho}}]{Yao2014}%
  \BibitemOpen
  \bibfield  {author} {\bibinfo {author} {\bibfnamefont {Y.~X.}\ \bibnamefont
  {Yao}}, \bibinfo {author} {\bibfnamefont {J.}~\bibnamefont {Liu}}, \bibinfo
  {author} {\bibfnamefont {C.~Z.}\ \bibnamefont {Wang}},\ and\ \bibinfo
  {author} {\bibfnamefont {K.~M.}\ \bibnamefont {Ho}},\ }\href
  {https://doi.org/10.1103/PhysRevB.89.045131} {\bibfield  {journal} {\bibinfo
  {journal} {Phys. Rev. B}\ }\textbf {\bibinfo {volume} {89}},\ \bibinfo
  {pages} {045131} (\bibinfo {year} {2014})}\BibitemShut {NoStop}%
\bibitem [{\citenamefont {Yao}\ \emph {et~al.}(2015)\citenamefont {Yao},
  \citenamefont {Liu}, \citenamefont {Liu}, \citenamefont {Lu}, \citenamefont
  {Wang},\ and\ \citenamefont {Ho}}]{Yao2015}%
  \BibitemOpen
  \bibfield  {author} {\bibinfo {author} {\bibfnamefont {Y.~X.}\ \bibnamefont
  {Yao}}, \bibinfo {author} {\bibfnamefont {J.}~\bibnamefont {Liu}}, \bibinfo
  {author} {\bibfnamefont {C.}~\bibnamefont {Liu}}, \bibinfo {author}
  {\bibfnamefont {W.~C.}\ \bibnamefont {Lu}}, \bibinfo {author} {\bibfnamefont
  {C.~Z.}\ \bibnamefont {Wang}},\ and\ \bibinfo {author} {\bibfnamefont
  {K.~M.}\ \bibnamefont {Ho}},\ }\href {https://doi.org/10.1038/srep13478}
  {\bibfield  {journal} {\bibinfo  {journal} {Sci. Rep.}\ }\textbf {\bibinfo
  {volume} {5}},\ \bibinfo {pages} {13478} (\bibinfo {year}
  {2015})}\BibitemShut {NoStop}%
\bibitem [{\citenamefont {Lanat\`a}\ \emph {et~al.}(2015)\citenamefont
  {Lanat\`a}, \citenamefont {Yao}, \citenamefont {Wang}, \citenamefont {Ho},\
  and\ \citenamefont {Kotliar}}]{lanata2015}%
  \BibitemOpen
  \bibfield  {author} {\bibinfo {author} {\bibfnamefont {N.}~\bibnamefont
  {Lanat\`a}}, \bibinfo {author} {\bibfnamefont {Y.}~\bibnamefont {Yao}},
  \bibinfo {author} {\bibfnamefont {C.-Z.}\ \bibnamefont {Wang}}, \bibinfo
  {author} {\bibfnamefont {K.-M.}\ \bibnamefont {Ho}},\ and\ \bibinfo {author}
  {\bibfnamefont {G.}~\bibnamefont {Kotliar}},\ }\href
  {https://doi.org/10.1103/PhysRevX.5.011008} {\bibfield  {journal} {\bibinfo
  {journal} {Phys. Rev. X}\ }\textbf {\bibinfo {volume} {5}},\ \bibinfo {pages}
  {011008} (\bibinfo {year} {2015})}\BibitemShut {NoStop}%
\bibitem [{\citenamefont {Fabrizio}(2017)}]{fabrizio2017}%
  \BibitemOpen
  \bibfield  {author} {\bibinfo {author} {\bibfnamefont {M.}~\bibnamefont
  {Fabrizio}},\ }\href {https://doi.org/10.1103/PhysRevB.95.075156} {\bibfield
  {journal} {\bibinfo  {journal} {Phys. Rev. B}\ }\textbf {\bibinfo {volume}
  {95}},\ \bibinfo {pages} {075156} (\bibinfo {year} {2017})}\BibitemShut
  {NoStop}%
\bibitem [{\citenamefont {Giuli}\ \emph
  {et~al.}(2025{\natexlab{b}})\citenamefont {Giuli}, \citenamefont
  {Mejuto-Zaera},\ and\ \citenamefont {Capone}}]{Giuli2025}%
  \BibitemOpen
  \bibfield  {author} {\bibinfo {author} {\bibfnamefont {S.}~\bibnamefont
  {Giuli}}, \bibinfo {author} {\bibfnamefont {C.}~\bibnamefont
  {Mejuto-Zaera}},\ and\ \bibinfo {author} {\bibfnamefont {M.}~\bibnamefont
  {Capone}},\ }\href {https://doi.org/10.1103/PhysRevB.111.L020401} {\bibfield
  {journal} {\bibinfo  {journal} {Phys. Rev. B}\ }\textbf {\bibinfo {volume}
  {111}},\ \bibinfo {pages} {L020401} (\bibinfo {year}
  {2025}{\natexlab{b}})}\BibitemShut {NoStop}%
\bibitem [{\citenamefont {Zhang}\ \emph {et~al.}(1988)\citenamefont {Zhang},
  \citenamefont {Gros}, \citenamefont {Rice},\ and\ \citenamefont
  {Shiba}}]{ZhangGrosRiceShiba1988}%
  \BibitemOpen
  \bibfield  {author} {\bibinfo {author} {\bibfnamefont {F.~C.}\ \bibnamefont
  {Zhang}}, \bibinfo {author} {\bibfnamefont {C.}~\bibnamefont {Gros}},
  \bibinfo {author} {\bibfnamefont {T.~M.}\ \bibnamefont {Rice}},\ and\
  \bibinfo {author} {\bibfnamefont {H.}~\bibnamefont {Shiba}},\ }\href
  {https://doi.org/10.1088/0953-2048/1/1/009} {\bibfield  {journal} {\bibinfo
  {journal} {Supercond. Sci. Technol.}\ }\textbf {\bibinfo {volume} {1}},\
  \bibinfo {pages} {36} (\bibinfo {year} {1988})}\BibitemShut {NoStop}%
\bibitem [{\citenamefont {Affleck}\ and\ \citenamefont
  {Marston}(1988)}]{Affleck-Marston-flux-PRB1988}%
  \BibitemOpen
  \bibfield  {author} {\bibinfo {author} {\bibfnamefont {I.}~\bibnamefont
  {Affleck}}\ and\ \bibinfo {author} {\bibfnamefont {J.~B.}\ \bibnamefont
  {Marston}},\ }\href {https://doi.org/10.1103/PhysRevB.37.3774} {\bibfield
  {journal} {\bibinfo  {journal} {Phys. Rev. B}\ }\textbf {\bibinfo {volume}
  {37}},\ \bibinfo {pages} {3774} (\bibinfo {year} {1988})}\BibitemShut
  {NoStop}%
\bibitem [{\citenamefont {Affleck}\ \emph {et~al.}(1988)\citenamefont
  {Affleck}, \citenamefont {Zou}, \citenamefont {Hsu},\ and\ \citenamefont
  {Anderson}}]{Affleck-flux-PRB1988}%
  \BibitemOpen
  \bibfield  {author} {\bibinfo {author} {\bibfnamefont {I.}~\bibnamefont
  {Affleck}}, \bibinfo {author} {\bibfnamefont {Z.}~\bibnamefont {Zou}},
  \bibinfo {author} {\bibfnamefont {T.}~\bibnamefont {Hsu}},\ and\ \bibinfo
  {author} {\bibfnamefont {P.~W.}\ \bibnamefont {Anderson}},\ }\href
  {https://doi.org/10.1103/PhysRevB.38.745} {\bibfield  {journal} {\bibinfo
  {journal} {Phys. Rev. B}\ }\textbf {\bibinfo {volume} {38}},\ \bibinfo
  {pages} {745} (\bibinfo {year} {1988})}\BibitemShut {NoStop}%
\bibitem [{\citenamefont {Sandri}\ \emph {et~al.}(2013)\citenamefont {Sandri},
  \citenamefont {Capone},\ and\ \citenamefont {Fabrizio}}]{Sandri2013}%
  \BibitemOpen
  \bibfield  {author} {\bibinfo {author} {\bibfnamefont {M.}~\bibnamefont
  {Sandri}}, \bibinfo {author} {\bibfnamefont {M.}~\bibnamefont {Capone}},\
  and\ \bibinfo {author} {\bibfnamefont {M.}~\bibnamefont {Fabrizio}},\ }\href
  {https://doi.org/10.1103/PhysRevB.87.205108} {\bibfield  {journal} {\bibinfo
  {journal} {Phys. Rev. B}\ }\textbf {\bibinfo {volume} {87}},\ \bibinfo
  {pages} {205108} (\bibinfo {year} {2013})}\BibitemShut {NoStop}%
\bibitem [{\citenamefont {Mila}\ and\ \citenamefont {Rice}(1989)}]{MilaRice1989}%
  \BibitemOpen
  \bibfield  {author} {\bibinfo {author} {\bibfnamefont {F.}~\bibnamefont
  {Mila}}\ and\ \bibinfo {author} {\bibfnamefont {T.~M.}\ \bibnamefont
  {Rice}},\ }\href {https://doi.org/10.1016/0921-4534(89)90286-4} {\bibfield
  {journal} {\bibinfo  {journal} {Physica C}\ }\textbf {\bibinfo {volume}
  {157}},\ \bibinfo {pages} {561} (\bibinfo {year} {1989})}\BibitemShut {NoStop}%
\bibitem [{\citenamefont {Alloul}\ \emph {et~al.}(1989)\citenamefont
  {Alloul}, \citenamefont {Ohno},\ and\ \citenamefont {Mendels}}]{Alloul1989}%
  \BibitemOpen
  \bibfield  {author} {\bibinfo {author} {\bibfnamefont {H.}~\bibnamefont
  {Alloul}}, \bibinfo {author} {\bibfnamefont {T.}~\bibnamefont {Ohno}},\ and\
  \bibinfo {author} {\bibfnamefont {P.}~\bibnamefont {Mendels}},\ }\href
  {https://doi.org/10.1103/PhysRevLett.63.1700} {\bibfield  {journal}
  {\bibinfo  {journal} {Phys. Rev. Lett.}\ }\textbf {\bibinfo {volume} {63}},\
  \bibinfo {pages} {1700} (\bibinfo {year} {1989})}\BibitemShut {NoStop}%
\bibitem [{\citenamefont {Takigawa}\ \emph {et~al.}(1991)\citenamefont
  {Takigawa}, \citenamefont {Reyes}, \citenamefont {Hammel}, \citenamefont
  {Thompson}, \citenamefont {Heffner}, \citenamefont {Fisk},\ and\ \citenamefont
  {Ott}}]{Takigawa1991}%
  \BibitemOpen
  \bibfield  {author} {\bibinfo {author} {\bibfnamefont {M.}~\bibnamefont
  {Takigawa}}, \bibinfo {author} {\bibfnamefont {A.~P.}\ \bibnamefont
  {Reyes}}, \bibinfo {author} {\bibfnamefont {P.~C.}\ \bibnamefont {Hammel}},
  \bibinfo {author} {\bibfnamefont {J.~D.}\ \bibnamefont {Thompson}}, \bibinfo
  {author} {\bibfnamefont {R.~H.}\ \bibnamefont {Heffner}}, \bibinfo {author}
  {\bibfnamefont {Z.}~\bibnamefont {Fisk}},\ and\ \bibinfo {author}
  {\bibfnamefont {K.~C.}\ \bibnamefont {Ott}},\ }\href
  {https://doi.org/10.1103/PhysRevB.43.247} {\bibfield  {journal} {\bibinfo
  {journal} {Phys. Rev. B}\ }\textbf {\bibinfo {volume} {43}},\ \bibinfo
  {pages} {247} (\bibinfo {year} {1991})}\BibitemShut {NoStop}%
\bibitem [{\citenamefont {Moriya}(1963)}]{Moriya1963}%
  \BibitemOpen
  \bibfield  {author} {\bibinfo {author} {\bibfnamefont {T.}~\bibnamefont
  {Moriya}},\ }\href {https://doi.org/10.1143/JPSJ.18.516} {\bibfield
  {journal} {\bibinfo  {journal} {J. Phys. Soc. Jpn.}\ }\textbf {\bibinfo
  {volume} {18}},\ \bibinfo {pages} {516} (\bibinfo {year} {1963})}\BibitemShut
  {NoStop}%
\bibitem [{\citenamefont {Millis}\ \emph {et~al.}(1990)\citenamefont {Millis},
  \citenamefont {Monien},\ and\ \citenamefont {Pines}}]{MillisMonienPines1990}%
  \BibitemOpen
  \bibfield  {author} {\bibinfo {author} {\bibfnamefont {A.~J.}\ \bibnamefont
  {Millis}}, \bibinfo {author} {\bibfnamefont {H.}~\bibnamefont {Monien}},\
  and\ \bibinfo {author} {\bibfnamefont {D.}~\bibnamefont {Pines}},\ }\href
  {https://doi.org/10.1103/PhysRevB.42.167} {\bibfield  {journal} {\bibinfo
  {journal} {Phys. Rev. B}\ }\textbf {\bibinfo {volume} {42}},\ \bibinfo
  {pages} {167} (\bibinfo {year} {1990})}\BibitemShut {NoStop}%
\bibitem [{\citenamefont {Pavarini}\ \emph {et~al.}(2001)\citenamefont
  {Pavarini}, \citenamefont {Dasgupta}, \citenamefont {Saha-Dasgupta},
  \citenamefont {Jepsen},\ and\ \citenamefont {Andersen}}]{Pavarini2001}%
  \BibitemOpen
  \bibfield  {author} {\bibinfo {author} {\bibfnamefont {E.}~\bibnamefont
  {Pavarini}}, \bibinfo {author} {\bibfnamefont {I.}~\bibnamefont {Dasgupta}},
  \bibinfo {author} {\bibfnamefont {T.}~\bibnamefont {Saha-Dasgupta}}, \bibinfo
  {author} {\bibfnamefont {O.}~\bibnamefont {Jepsen}},\ and\ \bibinfo {author}
  {\bibfnamefont {O.~K.}\ \bibnamefont {Andersen}},\ }\href
  {https://doi.org/10.1103/PhysRevLett.87.047003} {\bibfield  {journal}
  {\bibinfo  {journal} {Phys. Rev. Lett.}\ }\textbf {\bibinfo {volume} {87}},\
  \bibinfo {pages} {047003} (\bibinfo {year} {2001})}\BibitemShut {NoStop}%
\bibitem [{\citenamefont {Shih}\ \emph {et~al.}(2004)\citenamefont {Shih},
  \citenamefont {Lee}, \citenamefont {Eder}, \citenamefont {Mou},\ and\
  \citenamefont {Chen}}]{Shih2004}%
  \BibitemOpen
  \bibfield  {author} {\bibinfo {author} {\bibfnamefont {C.~T.}\ \bibnamefont
  {Shih}}, \bibinfo {author} {\bibfnamefont {T.~K.}\ \bibnamefont {Lee}},
  \bibinfo {author} {\bibfnamefont {R.}~\bibnamefont {Eder}}, \bibinfo {author}
  {\bibfnamefont {C.-Y.}\ \bibnamefont {Mou}},\ and\ \bibinfo {author}
  {\bibfnamefont {Y.~C.}\ \bibnamefont {Chen}},\ }\href
  {https://doi.org/10.1103/PhysRevLett.92.227002} {\bibfield  {journal}
  {\bibinfo  {journal} {Phys. Rev. Lett.}\ }\textbf {\bibinfo {volume} {92}},\
  \bibinfo {pages} {227002} (\bibinfo {year} {2004})}\BibitemShut {NoStop}%
\bibitem [{\citenamefont {Zhou}\ \emph {et~al.}(2025)\citenamefont {Zhou},
  \citenamefont {Vinograd}, \citenamefont {Hirata}, \citenamefont {Wu},
  \citenamefont {Mayaffre}, \citenamefont {Kr\"amer}, \citenamefont {Hardy},
  \citenamefont {Liang}, \citenamefont {Bonn}, \citenamefont {Loew},
  \citenamefont {Porras}, \citenamefont {Keimer},\ and\ \citenamefont
  {Julien}}]{Zhou2025}%
  \BibitemOpen
  \bibfield  {author} {\bibinfo {author} {\bibfnamefont {R.}~\bibnamefont
  {Zhou}}, \bibinfo {author} {\bibfnamefont {I.}~\bibnamefont {Vinograd}},
  \bibinfo {author} {\bibfnamefont {M.}~\bibnamefont {Hirata}}, \bibinfo
  {author} {\bibfnamefont {T.}~\bibnamefont {Wu}}, \bibinfo {author}
  {\bibfnamefont {H.}~\bibnamefont {Mayaffre}}, \bibinfo {author}
  {\bibfnamefont {S.}~\bibnamefont {Kr\"amer}}, \bibinfo {author}
  {\bibfnamefont {W.~N.}\ \bibnamefont {Hardy}}, \bibinfo {author}
  {\bibfnamefont {R.}~\bibnamefont {Liang}}, \bibinfo {author} {\bibfnamefont
  {D.~A.}\ \bibnamefont {Bonn}}, \bibinfo {author} {\bibfnamefont
  {T.}~\bibnamefont {Loew}}, \bibinfo {author} {\bibfnamefont
  {J.}~\bibnamefont {Porras}}, \bibinfo {author} {\bibfnamefont
  {B.}~\bibnamefont {Keimer}},\ and\ \bibinfo {author} {\bibfnamefont
  {M.-H.}\ \bibnamefont {Julien}},\ }\href
  {https://doi.org/10.1038/s41567-024-02692-w} {\bibfield  {journal} {\bibinfo
  {journal} {Nat. Phys.}\ }\textbf {\bibinfo {volume} {21}},\ \bibinfo {pages}
  {97} (\bibinfo {year} {2025})}\BibitemShut {NoStop}%
\bibitem [{\citenamefont {Chen}\ \emph {et~al.}(2017)\citenamefont {Chen},
  \citenamefont {LeBlanc},\ and\ \citenamefont {Gull}}]{Chen2017}%
  \BibitemOpen
  \bibfield  {author} {\bibinfo {author} {\bibfnamefont {X.}~\bibnamefont
  {Chen}}, \bibinfo {author} {\bibfnamefont {J.~P.~F.}\ \bibnamefont
  {LeBlanc}},\ and\ \bibinfo {author} {\bibfnamefont {E.}~\bibnamefont
  {Gull}},\ }\href {https://doi.org/10.1038/ncomms14986} {\bibfield  {journal}
  {\bibinfo  {journal} {Nat. Commun.}\ }\textbf {\bibinfo {volume} {8}},\
  \bibinfo {pages} {14986} (\bibinfo {year} {2017})}\BibitemShut {NoStop}%
\bibitem [{\citenamefont {Sordi}\ \emph {et~al.}(2012)\citenamefont {Sordi},
  \citenamefont {S\'emon}, \citenamefont {Haule},\ and\ \citenamefont
  {Tremblay}}]{Sordi2012}%
  \BibitemOpen
  \bibfield  {author} {\bibinfo {author} {\bibfnamefont {G.}~\bibnamefont
  {Sordi}}, \bibinfo {author} {\bibfnamefont {P.}~\bibnamefont {S\'emon}},
  \bibinfo {author} {\bibfnamefont {K.}~\bibnamefont {Haule}},\ and\ \bibinfo
  {author} {\bibfnamefont {A.-M.~S.}\ \bibnamefont {Tremblay}},\ }\href
  {https://doi.org/10.1038/srep00547} {\bibfield  {journal} {\bibinfo
  {journal} {Sci. Rep.}\ }\textbf {\bibinfo {volume} {2}},\ \bibinfo {pages}
  {547} (\bibinfo {year} {2012})}\BibitemShut {NoStop}%
\bibitem [{\citenamefont {Macridin}\ \emph {et~al.}(2006)\citenamefont
  {Macridin}, \citenamefont {Jarrell}, \citenamefont {Maier}, \citenamefont
  {Kent},\ and\ \citenamefont {D'Azevedo}}]{Macridin2006}%
  \BibitemOpen
  \bibfield  {author} {\bibinfo {author} {\bibfnamefont {A.}~\bibnamefont
  {Macridin}}, \bibinfo {author} {\bibfnamefont {M.}~\bibnamefont {Jarrell}},
  \bibinfo {author} {\bibfnamefont {T.}~\bibnamefont {Maier}}, \bibinfo
  {author} {\bibfnamefont {P.~R.~C.}\ \bibnamefont {Kent}},\ and\ \bibinfo
  {author} {\bibfnamefont {E.}~\bibnamefont {D'Azevedo}},\ }\href
  {https://doi.org/10.1103/PhysRevLett.97.036401} {\bibfield  {journal}
  {\bibinfo  {journal} {Phys. Rev. Lett.}\ }\textbf {\bibinfo {volume} {97}},\
  \bibinfo {pages} {036401} (\bibinfo {year} {2006})}\BibitemShut {NoStop}%
\bibitem [{\citenamefont {Malcolms}\ \emph {et~al.}(2026)\citenamefont
  {Malcolms}, \citenamefont {Menke}, \citenamefont {Tseng}, \citenamefont
  {Jacob}, \citenamefont {Held}, \citenamefont {Hansmann},\ and\ \citenamefont
  {Sch\"afer}}]{Malcolms2026}%
  \BibitemOpen
  \bibfield  {author} {\bibinfo {author} {\bibfnamefont {M.~O.}\ \bibnamefont
  {Malcolms}}, \bibinfo {author} {\bibfnamefont {H.}~\bibnamefont {Menke}},
  \bibinfo {author} {\bibfnamefont {Y.-T.}\ \bibnamefont {Tseng}}, \bibinfo
  {author} {\bibfnamefont {E.}~\bibnamefont {Jacob}}, \bibinfo {author}
  {\bibfnamefont {K.}~\bibnamefont {Held}}, \bibinfo {author} {\bibfnamefont
  {P.}~\bibnamefont {Hansmann}},\ and\ \bibinfo {author} {\bibfnamefont
  {T.}~\bibnamefont {Sch\"afer}},\ }\href
  {https://doi.org/10.1038/s42005-026-02685-6} {\bibfield  {journal} {\bibinfo
  {journal} {Commun. Phys.}\ }\textbf {\bibinfo {volume} {9}},\ \bibinfo
  {pages} {179} (\bibinfo {year} {2026})}\BibitemShut {NoStop}%
\bibitem [{\citenamefont {Hettler}\ \emph {et~al.}(1998)\citenamefont
  {Hettler}, \citenamefont {Tahvildar-Zadeh}, \citenamefont {Jarrell},
  \citenamefont {Pruschke},\ and\ \citenamefont {Krishnamurthy}}]{Hettler1998}%
  \BibitemOpen
  \bibfield  {author} {\bibinfo {author} {\bibfnamefont {M.~H.}\ \bibnamefont
  {Hettler}}, \bibinfo {author} {\bibfnamefont {A.~N.}\ \bibnamefont
  {Tahvildar-Zadeh}}, \bibinfo {author} {\bibfnamefont {M.}~\bibnamefont
  {Jarrell}}, \bibinfo {author} {\bibfnamefont {T.}~\bibnamefont {Pruschke}},\
  and\ \bibinfo {author} {\bibfnamefont {H.~R.}\ \bibnamefont
  {Krishnamurthy}},\ }\href {https://doi.org/10.1103/PhysRevB.58.R7475}
  {\bibfield  {journal} {\bibinfo  {journal} {Phys. Rev. B}\ }\textbf {\bibinfo
  {volume} {58}},\ \bibinfo {pages} {R7475} (\bibinfo {year}
  {1998})}\BibitemShut {NoStop}%
\bibitem [{\citenamefont {Kotliar}\ \emph {et~al.}(2001)\citenamefont
  {Kotliar}, \citenamefont {Savrasov}, \citenamefont {P\'alsson},\ and\
  \citenamefont {Biroli}}]{Kotliar2001}%
  \BibitemOpen
  \bibfield  {author} {\bibinfo {author} {\bibfnamefont {G.}~\bibnamefont
  {Kotliar}}, \bibinfo {author} {\bibfnamefont {S.~Y.}\ \bibnamefont
  {Savrasov}}, \bibinfo {author} {\bibfnamefont {G.}~\bibnamefont
  {P\'alsson}},\ and\ \bibinfo {author} {\bibfnamefont {G.}~\bibnamefont
  {Biroli}},\ }\href {https://doi.org/10.1103/PhysRevLett.87.186401} {\bibfield
   {journal} {\bibinfo  {journal} {Phys. Rev. Lett.}\ }\textbf {\bibinfo
  {volume} {87}},\ \bibinfo {pages} {186401} (\bibinfo {year}
  {2001})}\BibitemShut {NoStop}%
\bibitem [{\citenamefont {Maier}\ \emph {et~al.}(2005)\citenamefont {Maier},
  \citenamefont {Jarrell}, \citenamefont {Pruschke},\ and\ \citenamefont
  {Hettler}}]{Maier2005}%
  \BibitemOpen
  \bibfield  {author} {\bibinfo {author} {\bibfnamefont {T.}~\bibnamefont
  {Maier}}, \bibinfo {author} {\bibfnamefont {M.}~\bibnamefont {Jarrell}},
  \bibinfo {author} {\bibfnamefont {T.}~\bibnamefont {Pruschke}},\ and\
  \bibinfo {author} {\bibfnamefont {M.~H.}\ \bibnamefont {Hettler}},\ }\href
  {https://doi.org/10.1103/RevModPhys.77.1027} {\bibfield  {journal} {\bibinfo
  {journal} {Rev. Mod. Phys.}\ }\textbf {\bibinfo {volume} {77}},\ \bibinfo
  {pages} {1027} (\bibinfo {year} {2005})}\BibitemShut {NoStop}%
\bibitem [{\citenamefont {Gros}(1988)}]{Gros1988}%
  \BibitemOpen
  \bibfield  {author} {\bibinfo {author} {\bibfnamefont {C.}~\bibnamefont
  {Gros}},\ }\href {https://doi.org/10.1103/PhysRevB.38.931} {\bibfield
  {journal} {\bibinfo  {journal} {Phys. Rev. B}\ }\textbf {\bibinfo {volume}
  {38}},\ \bibinfo {pages} {931} (\bibinfo {year} {1988})}\BibitemShut
  {NoStop}%
\bibitem [{\citenamefont {Becca}\ and\ \citenamefont
  {Sorella}(2017)}]{BeccaSorella2017}%
  \BibitemOpen
  \bibfield  {author} {\bibinfo {author} {\bibfnamefont {F.}~\bibnamefont
  {Becca}}\ and\ \bibinfo {author} {\bibfnamefont {S.}~\bibnamefont
  {Sorella}},\ }\href {https://doi.org/10.1017/9781316417041} {\emph {\bibinfo
  {title} {Quantum Monte Carlo Approaches for Correlated Systems}}}\ (\bibinfo
  {publisher} {Cambridge University Press},\ \bibinfo {address} {Cambridge},\
  \bibinfo {year} {2017})\BibitemShut {NoStop}%
\bibitem [{\citenamefont {Rende}\ \emph {et~al.}(2026)\citenamefont {Rende},
  \citenamefont {Nikolaenko}, \citenamefont {Viteritti}, \citenamefont
  {Sachdev},\ and\ \citenamefont {Zhang}}]{Rende2026}%
  \BibitemOpen
  \bibfield  {author} {\bibinfo {author} {\bibfnamefont {R.}~\bibnamefont
  {Rende}}, \bibinfo {author} {\bibfnamefont {A.}~\bibnamefont {Nikolaenko}},
  \bibinfo {author} {\bibfnamefont {L.~L.}\ \bibnamefont {Viteritti}}, \bibinfo
  {author} {\bibfnamefont {S.}~\bibnamefont {Sachdev}},\ and\ \bibinfo {author}
  {\bibfnamefont {Y.-H.}\ \bibnamefont {Zhang}}\ }\href
  {https://doi.org/10.48550/arXiv.2603.02316} {10.48550/arXiv.2603.02316}
  (\bibinfo {year} {2026}),\ \Eprint {https://arxiv.org/abs/2603.02316}
  {arXiv:2603.02316 [cond-mat.str-el]} \BibitemShut {NoStop}%
\bibitem [{\citenamefont {Dzero}\ \emph {et~al.}(2016)\citenamefont {Dzero},
  \citenamefont {Xia}, \citenamefont {Galitski},\ and\ \citenamefont
  {Coleman}}]{Dzero2016}%
  \BibitemOpen
  \bibfield  {author} {\bibinfo {author} {\bibfnamefont {M.}~\bibnamefont
  {Dzero}}, \bibinfo {author} {\bibfnamefont {J.}~\bibnamefont {Xia}}, \bibinfo
  {author} {\bibfnamefont {V.}~\bibnamefont {Galitski}},\ and\ \bibinfo
  {author} {\bibfnamefont {P.}~\bibnamefont {Coleman}},\ }\href
  {https://doi.org/10.1146/annurev-conmatphys-031214-014749} {\bibfield
  {journal} {\bibinfo  {journal} {Annu. Rev. Condens. Matter Phys.}\ }\textbf
  {\bibinfo {volume} {7}},\ \bibinfo {pages} {249} (\bibinfo {year}
  {2016})}\BibitemShut {NoStop}%
\bibitem [{\citenamefont {Tan}\ \emph {et~al.}(2015)\citenamefont {Tan},
  \citenamefont {Hsu}, \citenamefont {Zeng}, \citenamefont {Hatnean},
  \citenamefont {Harrison}, \citenamefont {Zhu}, \citenamefont {Hartstein},
  \citenamefont {Kiourlappou}, \citenamefont {Srivastava}, \citenamefont
  {Johannes}, \citenamefont {Murphy}, \citenamefont {Park}, \citenamefont
  {Balicas}, \citenamefont {Lonzarich}, \citenamefont {Balakrishnan},\ and\
  \citenamefont {Sebastian}}]{Tan2015}%
  \BibitemOpen
  \bibfield  {author} {\bibinfo {author} {\bibfnamefont {B.~S.}\ \bibnamefont
  {Tan}}, \bibinfo {author} {\bibfnamefont {Y.-T.}\ \bibnamefont {Hsu}},
  \bibinfo {author} {\bibfnamefont {B.}~\bibnamefont {Zeng}}, \bibinfo {author}
  {\bibfnamefont {M.~C.}\ \bibnamefont {Hatnean}}, \bibinfo {author}
  {\bibfnamefont {N.}~\bibnamefont {Harrison}}, \bibinfo {author}
  {\bibfnamefont {Z.}~\bibnamefont {Zhu}}, \bibinfo {author} {\bibfnamefont
  {M.}~\bibnamefont {Hartstein}}, \bibinfo {author} {\bibfnamefont
  {M.}~\bibnamefont {Kiourlappou}}, \bibinfo {author} {\bibfnamefont
  {A.}~\bibnamefont {Srivastava}}, \bibinfo {author} {\bibfnamefont {M.~D.}\
  \bibnamefont {Johannes}}, \bibinfo {author} {\bibfnamefont {T.~P.}\
  \bibnamefont {Murphy}}, \bibinfo {author} {\bibfnamefont {J.-H.}\
  \bibnamefont {Park}}, \bibinfo {author} {\bibfnamefont {L.}~\bibnamefont
  {Balicas}}, \bibinfo {author} {\bibfnamefont {G.~G.}\ \bibnamefont
  {Lonzarich}}, \bibinfo {author} {\bibfnamefont {G.}~\bibnamefont
  {Balakrishnan}},\ and\ \bibinfo {author} {\bibfnamefont {S.~E.}\ \bibnamefont
  {Sebastian}},\ }\href {https://doi.org/10.1126/science.aaa7974} {\bibfield
  {journal} {\bibinfo  {journal} {Science}\ }\textbf {\bibinfo {volume}
  {349}},\ \bibinfo {pages} {287} (\bibinfo {year} {2015})}\BibitemShut
  {NoStop}%
\bibitem [{\citenamefont {Hartstein}\ \emph {et~al.}(2018)\citenamefont
  {Hartstein}, \citenamefont {Toews}, \citenamefont {Hsu}, \citenamefont
  {Zeng}, \citenamefont {Chen}, \citenamefont {Hatnean}, \citenamefont {Zhang},
  \citenamefont {Nakamura}, \citenamefont {Padgett}, \citenamefont
  {Rodway-Gant}, \citenamefont {Berk}, \citenamefont {Kingston}, \citenamefont
  {Zhang}, \citenamefont {Chan}, \citenamefont {Yamashita}, \citenamefont
  {Sakakibara}, \citenamefont {Takano}, \citenamefont {Park}, \citenamefont
  {Balicas}, \citenamefont {Harrison}, \citenamefont {Shitsevalova},
  \citenamefont {Balakrishnan}, \citenamefont {Lonzarich}, \citenamefont
  {Hill}, \citenamefont {Sutherland},\ and\ \citenamefont
  {Sebastian}}]{Hartstein2018}%
  \BibitemOpen
  \bibfield  {author} {\bibinfo {author} {\bibfnamefont {M.}~\bibnamefont
  {Hartstein}}, \bibinfo {author} {\bibfnamefont {W.~H.}\ \bibnamefont
  {Toews}}, \bibinfo {author} {\bibfnamefont {Y.-T.}\ \bibnamefont {Hsu}},
  \bibinfo {author} {\bibfnamefont {B.}~\bibnamefont {Zeng}}, \bibinfo {author}
  {\bibfnamefont {X.}~\bibnamefont {Chen}}, \bibinfo {author} {\bibfnamefont
  {M.~C.}\ \bibnamefont {Hatnean}}, \bibinfo {author} {\bibfnamefont {Q.~R.}\
  \bibnamefont {Zhang}}, \bibinfo {author} {\bibfnamefont {S.}~\bibnamefont
  {Nakamura}}, \bibinfo {author} {\bibfnamefont {A.~S.}\ \bibnamefont
  {Padgett}}, \bibinfo {author} {\bibfnamefont {G.}~\bibnamefont
  {Rodway-Gant}}, \bibinfo {author} {\bibfnamefont {J.}~\bibnamefont {Berk}},
  \bibinfo {author} {\bibfnamefont {M.~K.}\ \bibnamefont {Kingston}}, \bibinfo
  {author} {\bibfnamefont {G.~H.}\ \bibnamefont {Zhang}}, \bibinfo {author}
  {\bibfnamefont {M.~K.}\ \bibnamefont {Chan}}, \bibinfo {author}
  {\bibfnamefont {S.}~\bibnamefont {Yamashita}}, \bibinfo {author}
  {\bibfnamefont {T.}~\bibnamefont {Sakakibara}}, \bibinfo {author}
  {\bibfnamefont {Y.}~\bibnamefont {Takano}}, \bibinfo {author} {\bibfnamefont
  {J.-H.}\ \bibnamefont {Park}}, \bibinfo {author} {\bibfnamefont
  {L.}~\bibnamefont {Balicas}}, \bibinfo {author} {\bibfnamefont
  {N.}~\bibnamefont {Harrison}}, \bibinfo {author} {\bibfnamefont
  {N.}~\bibnamefont {Shitsevalova}}, \bibinfo {author} {\bibfnamefont
  {G.}~\bibnamefont {Balakrishnan}}, \bibinfo {author} {\bibfnamefont {G.~G.}\
  \bibnamefont {Lonzarich}}, \bibinfo {author} {\bibfnamefont {R.~W.}\
  \bibnamefont {Hill}}, \bibinfo {author} {\bibfnamefont {M.}~\bibnamefont
  {Sutherland}},\ and\ \bibinfo {author} {\bibfnamefont {S.~E.}\ \bibnamefont
  {Sebastian}},\ }\href {https://doi.org/10.1038/nphys4295} {\bibfield
  {journal} {\bibinfo  {journal} {Nat. Phys.}\ }\textbf {\bibinfo {volume}
  {14}},\ \bibinfo {pages} {166} (\bibinfo {year} {2018})}\BibitemShut
  {NoStop}%
\bibitem [{\citenamefont {Hiai}\ and\ \citenamefont
  {Petz}(1993)}]{HiaiPetz1993}%
  \BibitemOpen
  \bibfield  {author} {\bibinfo {author} {\bibfnamefont {F.}~\bibnamefont
  {Hiai}}\ and\ \bibinfo {author} {\bibfnamefont {D.}~\bibnamefont {Petz}},\
  }\href {https://doi.org/10.1016/0024-3795(93)90029-N} {\bibfield  {journal}
  {\bibinfo  {journal} {Linear Algebra Appl.}\ }\textbf {\bibinfo {volume}
  {181}},\ \bibinfo {pages} {153} (\bibinfo {year} {1993})}\BibitemShut
  {NoStop}%
\end{thebibliography}
\end{document}